\documentclass[aps,pre,reprint,10pt]{revtex4-1}

\usepackage{amsmath,amssymb,bbm,mathtools}

\usepackage{graphicx}
\usepackage{bm}

\newcommand{\one}{\mathbbm{1}} 

\renewcommand{\Re}{\operatorname{Re}}\newcommand\RE\Re 
\renewcommand{\Im}{\operatorname{Im}}\newcommand\IM\Im 

\newcommand{\nn}{\nonumber} 

\DeclareMathOperator{\Ort}{O}
\DeclareMathOperator{\USp}{USp}
\DeclareMathOperator{\U}{U}

\newcommand{\eins}{\one}

\newcommand{\jacobi}{J}
\newcommand{\gauss}{G}

\renewcommand{\H}{\mathbb H} %

\let\olddet\det
\renewcommand{\det}{\olddet\nolimits} 

\renewcommand{\phi}{\varphi} 
\renewcommand{\epsilon}{\varepsilon} 

\DeclareMathOperator{\tr}{Tr} 
\DeclareMathOperator*{\pf}{Pf} 
\DeclareMathOperator*{\diag}{diag} 
\DeclareMathOperator{\sign}{sign} 

\DeclarePairedDelimiter{\abs}{\lvert}{\rvert} 

\newcommand{\MeijerG}[8][\bigg]{G^{{ #2 },\,{ #3 }}_{{ #4 },\,{ #5 }} #1( \begin{matrix} #6 \\ #7 \end{matrix}\, #1\vert\, #8 #1)}

\begin{document}

\title[Product of rectangular Ginibre matrices]{Weak Commutation Relations\\ and Eigenvalue Statistics for Products of Rectangular Random Matrices}
\author{Jesper R. Ipsen}
\email{jipsen@physik.uni-bielefeld.de}
\author{Mario Kieburg}
\email{mario.kieburg@uni-bielefeld.de}
\affiliation{Department of Physics, Bielefeld University, Postfach 100131, D-33501 Bielefeld, Germany}

\date{\today}

\begin{abstract}
We study the joint probability density of the eigenvalues of a product of rectangular real, complex or quaternion random matrices in a unified way. The random matrices are distributed according to arbitrary probability densities, whose only restriction is the invariance under left and right multiplication by orthogonal, unitary or unitary symplectic matrices, respectively. We show that a product of rectangular matrices is statistically equivalent to a product of square matrices. Hereby we prove a weak commutation relation of the random matrices at finite matrix sizes, which previously have been discussed for infinite matrix size. Moreover we derive the joint probability densities of the eigenvalues. To illustrate our results we apply them to a product of random matrices drawn from Ginibre ensembles and Jacobi ensembles as well as a mixed version thereof. For these weights we show that  the product of complex random matrices yield a determinantal point process, while the real and quaternion matrix ensembles 
correspond to Pfaffian point processes. Our results are visualized by numerical simulations. Furthermore, we present an application to a transport on a closed, disordered chain coupled to a particle bath.
\end{abstract}

\maketitle

\section{Introduction}\label{intro}

Recently, products of random matrices have experienced a revival due to new mathematical insights about the statistics of the eigen- and singular values for finite as well as infinite matrix dimensions. Recent progress in the field has made it possible to study a product of an arbitrary number of random matrices of arbitrary size for certain matrix ensembles. The fact that the number of matrices and their size can be chosen freely, allows discussions of various limits. This includes macroscopic as well as microscopic structures for infinite matrix dimension, but also the limit where the number of matrices goes to infinite is available. Analogous to the study of individual random matrices, products of random matrices show a rich mathematical structure and various limits have revealed new universality classes, which are important in the physical sciences as well as in mathematics and beyond.

Products of random matrices have been applied to a broad spectrum of the physical sciences. To name only a few of them: Transport in disordered and chaotic systems~\cite{Productsrandom}, matrix-valued diffusions~\cite{2002PhRvE..66f6124J,Gudowska}, quantum chromodynamics at finite chemical potential~\cite{Osborn,Akemann2007}, Yang--Mills theory~\cite{YangMill}, and percolation theory (introduction of \cite{Mehtabook}). Furthermore, results about products of random matrices have been applied to fields beyond physics, such as the study of wireless telecommunication~\cite{TulinoVer} and finance~\cite{BouchaudLalMic} as well as directions within mathematics, e.g. free probability~\cite{Speicher}. In this work an example of chaotic transport (cf. sec.~\ref{sec1}) will be our main motivation, but we discuss the mathematical structure in a completely general setting. Hence the results presented in this paper can be directly applied to other situations as well.

Note that, even though certain symmetries of random matrices might be conserved under matrix multiplication (such as unitarity), in general a product matrix possesses less symmetry than the individual matrices. In particular, a product of Hermitian matrices will not generally be Hermitian itself, and the eigenvalues will spread into the complex plane. This loss of symmetry has let to a particular interest in products of non-Hermitian matrices, especially drawn from Gaussian ensembles~\cite{AkemannBur2012,AkemannStr2012,Ipsen2013,AkemannKieWei2013,AkemannKieIps2013}. These random matrix ensembles are also known as Ginibre ensembles or Wishart ensembles. The discussion of products of Ginibre matrices at finite matrix dimension can be considered as an extension of previous results related to the product of two matrices motivated by applications to quantum chromodynamics at finite chemical potential~\cite{Osborn,QCD,KanzieperSin}. In our work we will go beyond the restriction to Ginibre 
ensembles and study a set of general weights only 
restricted by the invariance under left- and right-multiplication of unitary matrices. Furthermore, we will discuss all three Dyson classes in a unified way. We illustrate the underlying structure with the two particular ensembles of Ginibre and  Jacobi (truncated unitary) matrices. These matrices are directly related to a transport on a closed one-dimensional chaotic chain in an environment as it is shown in Sec.~\ref{sec1}.

In Sec.~\ref{sec2} we will show that products of random matrices invariant under left- and right-multiplication of unitary matrices satisfy a weak commutation relation. This commutation relation holds even for finite matrix dimension and not only for infinite dimension as discussed in \cite{Burda2013}. Moreover it has important physical consequences. In the specific example of a closed one-dimensional chaotic chain in an environment it reflects the invariance under reordering of potential wells as long as we do not consider cross correlations. Note that the weak commutation relation presented here is completely general and represents a general physical property of reordering invariance. For instance the same mechanism implies that communication channels with progressive scattering in wireless telecommunication are invariant under reordering of the clusters of scatterers~\cite{AkemannKieIps2013}. A similar commutation relation has previously been discussed in the context of disordered wires with obstacles~\
cite{BM:1994}.

In Secs.~\ref{sec3} and~\ref{sec4} we explicitly discuss two explicit realizations of matrix ensembles, namely products of Ginibre and Jacobi matrices, as well as an intermix of these and apply this result to derive an expression of the Lyapunov exponent for the disordered chain proposed in Sec.~\ref{sec1}. Section~\ref{sec5} is devoted to conclusions and outlook, while some technical details are presented in the appendix.

\section[Motivation]{Motivation:\\Closed one-dimensional chaotic chain}\label{sec1}

In this section we consider a model of a unitary evolution. It differs from a similar quantum evolution discussed in Ref.~\cite{2002PhRvE..66f6124J,JanikWiecz} by the idea that there is a coupling to an environment. Thus it is more in the spirit of one-dimensional quantum transport in a disordered system~\cite{Schmidt1957,Comtet2010,Comtet2013}. The unitary evolution matrix is taken to be time independent, hence we do not model a diffusive system but a chaotic quantum system, which scatters the particles into an environment. Here the environment is also modelled as a chaotic quantum system. Due to the coupling to the environment the evolution operator acts non-unitarily on the studied subsystem. 

We consider a chain of $M$ Hilbert-spaces of dimension $N_1,\ldots,N_M$ arranged along a ring. The Hilbert-spaces can be constructed by isolated potential wells. The particles in these wells can jump from one well to a consecutive one only in one direction and cannot stay in a well. Thus we have a totally asymmetric quantum transport. The system is constructed such that each well is coupled to a joint particle bath (environment) which can absorb the particles while the total amount of particles (on the chain and in the bath) is fixed. Moreover the Hamiltonian shall be time independent such that the transfer matrix is given by the unitary matrix
\begin{equation}\label{evolution}
 U=\left(\begin{array}{cc} X\widehat{T} & V_1 \\ V_2 & V_3\end{array}\right).
\end{equation}
The translation matrix, $\widehat T$, is defined as
\begin{equation}
\label{translation}
 \widehat{T}=\left[\begin{array}{cccccc} 0 & \eins_{N_M} & 0 & & \cdots & 0 \\  & 0 & \eins_{N_1} & 0 & \cdots & 0 \\ \vdots &  & \ddots & \ddots &  & \vdots \\  &  &  &  &  &  \\ 0 &  &  \cdots & & 0 & \eins_{N_{M-2}} \\ \eins_{N_{M-1}} & 0 &  & \cdots &  & 0 \end{array}\right],
\end{equation}
such that particles can jump between consecutive wells in one direction only. The transport along the chain is determined by the block diagonal matrix
\begin{equation}\label{diagonalm}
X=\diag(X_1,\ldots,X_M)
\end{equation}
where the block matrices, $X_m$, are rectangular matrices with real ($\beta=1$), complex ($\beta=2$) or quaternion ($\beta=4$) entries chosen according to Dyson's three-fold way \cite{Dyson62,Mehtabook}, 
\begin{equation}
 X_j\in{\rm gl}_\beta(N_j,N_{j-1})\equiv
\left\{\begin{array}{cl} \mathbb{R}^{N_{j}\times N_{j-1}}, & \beta=1,\\ \mathbb{C}^{N_{j}\times N_{j-1}}, & \beta=2,\\ \mathbb{H}^{N_{j}\times N_{j-1}}, & \beta=4.\end{array}\right.
\end{equation}
We set $N_0=N_M$, since we consider a closed chain. The other three block matrices, $V_i$, in Eq.~\eqref{evolution} are chosen such that $U$ is orthogonal ($\beta=1$), unitary ($\beta=2$), or unitary symplectic ($\beta=4$), respectively.

Throughout this paper  we will use the standard $2\times 2$ matrix representation for the quaternion number field, $\H$, see e.g.~\cite{Mehtabook}.

Let $N_{\rm bath}$ be the dimension of the Hilbert space of the bath and $N_{\rm chain}=\sum_{j=1}^MN_j$ be the one of the chain. We assume that the jumping as well as the coupling to the bath is a stochastic process. Since we do not assume any additional symmetry breaking condition apart from the Dyson classification~\cite{Dyson62,Mehtabook} and the totally asymmetric process, the measure for $U$ is given by the Haar-measure of $\Ort(N_{\rm chain}+N_{\rm bath})$, $\U(N_{\rm chain}+N_{\rm bath})$ and $\USp(2N_{\rm chain}+2N_{\rm bath})$ for $\beta=1,2,4$, respectively.

For this purpose we briefly rederive the measure for a rectangular, truncated unitary matrix $X\in{\rm gl}_\beta(N_1,N_2)$ resulting from
\begin{equation}\label{unimatr}
 U=\left[\begin{array}{cc} X & V_1 \\ V_2 & V_3 \end{array}\right]\in\left\{\begin{array}{cl} \Ort(L), & \beta=1,\\ \U(L), & \beta=2,\\ {\rm USp}(2 L), & \beta=4,\end{array}\right.
\end{equation}
with $N_1,N_2<L$. In order to deal with all three Dyson classes in a unified way, we introduce the variable
\begin{equation}\label{gamma}
 \gamma=\left\{\begin{array}{cl} 1, & \beta=1,2,\\ 2, & \beta=4.\end{array}\right.
\end{equation}
The measure for the truncated matrix $X$ is given by~\cite{Zyczkowski2000}
\begin{equation}\label{truncmeasu}
  d\nu^{\rm(\jacobi)}(X)\propto d[X]\int d[V_1]d[V_2]d[V_3]\, \delta(UU^\dagger-\eins_{\gamma L}),
\end{equation}
where $d[X]$ and $d[V_j]$ denote the product of all independent differentials (there are $\beta$ real independent degrees of freedom per matrix entry). The Dirac $\delta$-function for matrices is defined by the product of all Dirac $\delta$-functions of all independent real entries. The Dirac $\delta$-function ensures that $U$ is orthogonal, unitary or unitary symplectic, respectively. It is straightforward to integrate out the irrelevant degrees of of freedom, $V_i$, which yields~\cite{Zyczkowski2000,AkemannKieBur2013}
\begin{multline}\label{Jacobimeas}
  d\nu^{\rm(\jacobi)}(X)\propto d[X]\Theta(\eins_{\gamma N_1}-XX^{\dagger})\,
{\det}^{\kappa}(\eins_{\gamma N_1}-XX^{\dagger}),
\end{multline}
which is known as the Jacobi ensemble~\cite{Mehtabook,Forresterbook,Oxfordbook} and is labelled by a superscript $\text{(J)}$. The Heaviside $\Theta$-function for matrices is equal to unity for positive definite matrices and zero otherwise; and the power of the determinant, $\kappa$, is a constant given by $\kappa=\beta(L-N_1-N_2+1-2/\beta)/(2\gamma)$. The measure~\eqref{Jacobimeas} plays an important role in rest of this paper.

The discussion of the truncated unitary matrix, discussed in precious paragraph, can immediately be applied to the case described by Eq.~\eqref{evolution}. Integrating over $V_i$ in Eq.~\eqref{evolution}, we find
\begin{equation}\label{projection}
  d\nu(X)\propto d[X]\Theta(\eins_{\gamma N_{\rm chain}}-XX^{\dagger})\,
{\det}^{\kappa}(\eins_{\gamma N_{\rm chain}}-XX^{\dagger})
\end{equation}
with $\kappa=\beta(N_{\rm bath}-N_{\rm chain}+1-2/\beta)/(2\gamma)$.
The size of the bath, $N_\text{bath}$, and the size of the chain, $N_\text{chain}$, are independent quantities and one might be interested in the limit where the size of the bath goes to infinite while the size of the chain is kept fixed. The matrix $X$ will be of order $1/\sqrt{N_{\rm bath}}$, hence we rescale $X\to X/\sqrt{N_{\rm bath}}$. Thus for $N_{\rm bath}\gg N_{\rm chain}$ the measure of $X$ equals a Gaussian distribution,
\begin{equation}\label{measurelimit}
d\nu\left(\frac{X}{\sqrt{N_{\rm bath}}}\right)\propto \exp\left[-\frac{\beta }{2\gamma}\tr XX^\dagger\right],
\end{equation}
which follows from taking the limit in Eq.~\eqref{projection} without further restrictions. In particular, we do not need the central limit theorem, since the degrees of freedom are independent of $N_\text{bath}$. As a consequence the sub-matrices, see Eq.~\eqref{diagonalm}, are also Gaussian distributed,
\begin{equation}
d\nu^\text{(\gauss)}(X_j)\propto\exp\left[-\frac{\beta }{2\gamma}\tr X_jX_j^\dagger\right].
\end{equation}
This is also known as the Ginibre ensemble \cite{Mehtabook,Forresterbook,Oxfordbook} and is denoted by a superscript $\text{(G)}$. Depending on how large each well is, compared to the environment, one can also consider a mixed product of Ginibre and Jacobi matrices.

The distributions~\eqref{projection} and \eqref{measurelimit} enable us to calculate the spectral statistics of $U$ in the sector of the Hilbert-space representing the chain. In particular, we can consider the spectral statistics of $X\widehat{T}$. Note that the eigenvalues of $X\widehat{T}$ are intimately related to $(X\widehat{T})^{M}$ or, equivalently, to the product matrix $X^{(M)}\equiv X_MX_{M-1}\cdots X_2X_1$. Also the Lyapunov exponents defined as the logarithm of either the eigenvalues of $X^{(M)}$ or of the singular values (eigenvalues of $X^{(M)}X^{(M)\,\dagger}$) are widely used, see for example Refs.~\cite{Newman1986,Comtet2010,Comtet2013}. They measure the difference of a vector transported once around the chain by the non-unitary evolution

\section{Equivalence of different products of random matrices}\label{sec2}

An important property of products of rectangular random matrices is their relation to products of identically sized square matrices with deformed weights. The deformations are essentially prefactors of determinants of the random matrices and induce a repulsion from the origin, see Refs.~\cite{Verbaarschot2011,Fischmann2012,Ipsen2013,AkemannKieIps2013,Adhikari2013} for particular examples of Gaussian weights. In this section we study this relation for all three Dyson classes in a unified way. We emphasize that we do not specify a particular probability density for the random matrices in this section. The only assumption we enforce on the independent weights is invariance under one of the three groups $\Ort(N)$, $\U(N)$, and $\USp(2N)$ for $\beta=1,2,4$, respectively.

We consider the product of $M$ rectangular matrices,
\begin{equation}\label{defproduct}
  X^{(M)}=X_MX_{M-1}\cdots X_2X_1,
\end{equation}
where the individual matrices $X_j\in{\rm gl}_\beta(N_j,N_{j-1})$ have real, complex or quaternion entries according to the Dyson index, $\beta$. The rank of the product matrix is equal to $N_{\min}\equiv\min_{j=0,\ldots,M}N_j$. The product matrix is distributed according to the independent weights of the individual matrices,
\begin{align}\label{defmeasure}
 d\nu_{P_1,P_2,\ldots,P_M}(X^{(M)})&=\prod_{j=1}^MP_j(X_j)d[X_j],\\
d[X_j]&\equiv\prod_{a=1}^{N_j}\prod_{b=1}^{N_{j-1}}\prod_{\alpha=1}^{\beta}dX^{(j,\alpha)}_{ab}, \nn
\end{align}
where the product over $\alpha$ runs over all real degrees of freedom of a single matrix entry.
The only assumptions about the weights, $P_j$, is the invariance under left and right rotations,
\begin{equation}\label{inv}
  P_j(X_j)=P_j(V X_jU),
\end{equation}
for all transformations $V\otimes U$ in
\begin{equation}
 \U_\beta(N_j,N_{j-1})  \equiv
\begin{cases} \Ort(N_j)\otimes\Ort(N_{j-1}), & \beta=1,\\ \U(N_j)\otimes\U(N_{j-1}), & \beta=2,\\ \USp(2N_j)\otimes\USp(2N_{j-1}), & \beta=4.\end{cases}
\end{equation}
These probability densities were referred to as isotropic weights in Ref.~\cite{Burda2010-11}. Particular examples are: Gaussian weights (Ginibre ensembles) \cite{Newman1986,Verbaarschot2011,AkemannBur2012,Ipsen2013,AkemannKieWei2013,AkemannKieIps2013, Kuijlaars2013,Adhikari2013,Forrester2013},
\begin{equation}
 d\nu^{\rm (\gauss)}(X^{(M)})=\prod_{j=1}^M P^{\rm (\gauss)}(X_j)
 \propto\prod_{j=1}^M \exp\left[-\tr X_jX_j^\dagger\right],
\label{Gaussmeas}
\end{equation}
and weights which are the induced Haar measure of truncated unitary matrices (Jacobi ensembles)~\cite{Zyczkowski2000,Khoruzhenko2010,Adhikari2013,AkemannKieBur2013}
\begin{align}\label{truncmeas}
&d\nu_{\kappa}^{\rm (\jacobi)}(X^{(M)})=\prod_{j=1}^M P_{\kappa_j}^{\rm (\jacobi)}(X_j)\\
&\propto\prod_{j=1}^M {\det}^{\kappa_j}(\eins_{\gamma N_j}-X_jX_j^\dagger)\Theta(\eins_{\gamma N_j}-X_jX_j^\dagger)d[X_j],\nn
\end{align}
where $\kappa_j+\beta \min\{N_j,N_{j-1}\}/2+(\beta-2)/2>0$, compare with Eq.~\eqref{Jacobimeas}. Note that we deal with all three Dyson indices in a unified way. Hence the number of real independent degrees of freedom of a single matrix entry, $X^{(j)}_{ab}$, is equal to the Dyson index, $\beta$.

As will be discussed in detail below, a product of rectangular random matrices, $X^{(M)}$, can be expressed in terms of a product of square $\gamma N_{\min}\times\gamma N_{\min}$ random matrices and two truncated unitary matrices,
\begin{equation}\label{newproduct}
 X^{(M)}=U_{\rm L}\widetilde{X}^{(M)}U_{\rm R}=U_{\rm L}\widetilde{X}_M\widetilde{X}_{M-1}\cdots \widetilde{X}_2\widetilde{X}_1U_{\rm R}.
\end{equation}
Here  $\widetilde{X}_j\in{\rm gl}_\beta(N_{\min},N_{\min})$ are square matrices, while the $\gamma N_M\times\gamma N_{\min}$ matrix $U_{\rm L}$ consists of the first $\gamma N_{\min}$ columns of an element in the coset
\begin{multline}\label{coset}
  \mathbb{G}_\beta(N_M,N_{\min})\equiv\\
\begin{cases}
{\rm O}(N_M)/[{\rm O}(N_{\min})\times{\rm O}(\nu_M)], & \beta=1,\\ 
{\rm U}(N_M)/[{\rm U}(N_{\min})\times{\rm U}(\nu_M)], & \beta=2,\\ 
{\rm USp}(2 N_M)/[{\rm USp}(2 N_{\min})\times{\rm USp}(2\nu_M)], & \beta=4.
\end{cases}
\end{multline}
Likewise, the $\gamma N_{\min}\times\gamma N_0$ matrix $U_{\rm R}$ is equal to the first $\gamma N_{\min}$ rows of an element in the coset $\mathbb{G}_\beta(N_0,N_{\min})$. In~\eqref{coset} we have introduced the notation $\nu_j\equiv N_j-N_{\min}$. Note that the product matrix $\widetilde X^{(M)}$ is a square matrix. It is immediate from Eq.~\eqref{newproduct} that the nonzero singular values of $\widetilde X^{(M)}$ is identical to those of the original product matrix, $X^{(M)}$. Furthermore, if $X^{(M)}$ is a square matrix, then it turns out that also the eigenvalues will agree, except for $N_0-N_{\min}$ additional eigenvalues, which are all equal to zero. We will refer to the square product matrix, $\widetilde X^{(M)}$, as the induced product matrix and it can be considered as a generalization of the induced ensemble discussed in Ref.~\cite{Fischmann2012}.

Our main goal in this section is to derive the measure for the random matrix ${X}^{(M)}$ or equivalently the induced measures for the matrices $\widetilde{X}_j$ and $U_{\rm L}$ and $U_{\rm R}$. Furthermore, we establish a weak commutation relation for the square matrices, $\widetilde{X}_j$.

In subsection~\ref{sec2.1}, we present the two simplest cases, namely when one of the ``end-points'' of the chain of dimensions encountered in the product matrix, $X^{(M)}$, is equal to $\gamma N_{\min}$, i.e. $N_0=N_{\min}$ or $N_M=N_{\min}$. In these cases the resulting measure for ${X}^{(M)}$ can be readily derived. There seems to be an ambiguity of the resulting measure on the level of the individual random matrix measures. However, we show that commutativity of square random matrices does not only hold in the large $N$-limit, see Ref.~\cite{Burda2013}, but also at finite matrix size, see Sec.~\ref{sec2.2}. See also Ref.~\cite{BM:1994} In Sec.~\ref{sec2.3} we also derive a weak commutation relation between Jacobi matrices and an arbitrary isotropic random matrix which marginally changes the original weights. In subsection~\ref{sec2.4}, we discuss the general setting of arbitrary dimensions $N_j$ of the rectangular matrices in an arbitrary order.

\subsection{Two simple cases}\label{sec2.1}

We will first consider the case where $N_0=N_{\min}$, which implies that $U_{\rm R}$ is equal to unity. The matrix $X_1$ can easily be rotated to the $\gamma N_{\min}\times \gamma N_{\min}$ matrix $\widetilde X_1$ by a unitary transformation $U_1\in\mathbb{G}_\beta(N_1,N_{\min})$ from the left, which gives a block structure
\begin{equation}\label{leftder1}
 X_1=U_1\left[\begin{array}{c} \widetilde{X}_1 \\ 0 \end{array}\right].
\end{equation}
Here $0$ represents a $\gamma\nu_1\times\gamma N_{\min}$ matrix with all entries equal to zero (recall that $\nu_j=N_j-N_{\min}$).
The unitary matrix $U_1$ can be absorbed due to the $\U_\beta(N_2,N_1)$ invariance of the measure $P_2$, cf.~\eqref{inv}. Since the matrix $U_1$ is distributed according to the Haar measure on the coset $\mathbb{G}_\beta(N_1,N_{\min})$, the integral over $U_1$ completely factorizes from the rest and yields a constant. The change of coordinates~\eqref{leftder1} yields the measure
\begin{equation}\label{newmeas1}
 \widetilde{P}_1^\text{(L)}(\widetilde{X}_1 )\propto{\det}^{\beta\nu_1/(2\gamma)}(\widetilde{X}_1\widetilde{X}_1^\dagger)P_1\left(\left[\begin{array}{c} \widetilde{X}_1 \\ 0 \end{array}\right]\right)
\end{equation}
for $\widetilde{X}_1$. The superscript (L) denotes that we decompose $X_1$ via a left block QR-decomposition, see Eq.~\eqref{leftder1}. The determinantal prefactor comes from the change of coordinates and enforces an additional repulsion of the singular values as well as eigenvalues from the origin, cf. Refs.~\cite{Verbaarschot2011,Fischmann2012,Ipsen2013,AkemannKieIps2013,Adhikari2013}.

The smaller dimension of $\widetilde{X}_1$ projects the dimension on the right side of $X_2$ down to $\gamma N_{\min}$ in the product matrix $X^{(M)}$. Hence, we can again perform the same procedure as before for $X_1$. We bring $X_2$ into a block structure using a left block QR-decomposition
\begin{equation}\label{leftder2}
 X_2=U_2\left[\begin{array}{c|c} \begin{array}{c} \widetilde{X}_2 \\ 0 \end{array} & X'_2\end{array}\right],
\end{equation}
hence $X_2$ is decomposed into a unitary matrix times a block matrix consisting of two rectangular blocks.
The unitary matrix $U_2\in\mathbb{G}_\beta(N_2,N_{\min})$ is absorbed in the measure $P_3$. The integration over the rectangular matrix $X'_2\in{\rm gl}_\beta(N_2,\nu_1)$ is comprised in the definition of the reduced measure
\begin{equation}\label{newmeas2}
 \widetilde{P}_2^\text{(L)}(\widetilde{X}_2 )\propto{\det}^{\beta\nu_2/(2\gamma)}(\widetilde{X}_2\widetilde{X}_2^\dagger)\!\int\! d[X'_2] P_2\left(\left[\begin{array}{c|c} \begin{array}{c} \widetilde{X}_2 \\ 0 \end{array} & X'_2\end{array}\right]\right).
\end{equation}
Again the determinantal prefactor is the result of the degrees of freedom decoupling via the unitary matrix $U_2$. We repeat the same procedure for $X_3,X_4\ldots X_{M}$, and the new measures for the matrices $\widetilde{X}_j$, $2\leq j\leq M$, defined by the choice of coordinates
\begin{equation}\label{leftder3}
 X_j=U_j\left[\begin{array}{c|c} \begin{array}{c} \widetilde{X}_j \\ 0 \end{array} & X'_j\end{array}\right]
\end{equation}
are up to normalization constants
\begin{equation}\label{newmeas3}
 \widetilde{P}_j^\text{(L)}(\widetilde{X}_j )\propto{\det}^{\beta\nu_j/(2\gamma)}(\widetilde{X}_j\widetilde{X}_j^\dagger)\!\int\! d[X'_j] P_j\left(\left[\begin{array}{c|c} \begin{array}{c} \widetilde{X}_j \\ 0 \end{array} & X'_j\end{array}\right]\right),
 \end{equation}
where we integrate over the rectangular matrix $X'_j\in{\rm gl}_\beta(N_j,\nu_{j-1})$. Apart from $U_M\in\mathbb{G}_\beta(N_M,N_{\min})$ all unitary matrices can be absorbed in the measures $P_j$ because of their group invariance. Due to projection, $U_{\rm L}$ is a matrix consisting of the first $\gamma N_{\min}$ rows of the unitary matrix $U_M$. Thus the measure of the product matrix, ${X}^{(M)}$ is given by
\begin{equation}\label{measuregen1}
 d\nu_{\widetilde{P}_1^\text{(L)}\cdots\widetilde{P}_M^\text{(L)}}({X}^{(M)})= d\mu(U_{\rm L})\prod_{j=1}^M\widetilde{P}_j^\text{(L)}( \widetilde{X}_j)d[\widetilde{X}_j].
\end{equation}
Here $d\mu(U)$ denotes the Haar measure on $\mathbb{G}_\beta(N_M,N_{\min})$. Note that also the new weights, $\widetilde{P}_j^\text{(L)}$, are invariant under left and right multiplication of group elements in $\Ort(N_{\min})$, $\U(N_{\min})$, and $\USp(2N_{\min})$ for $\beta=1,2,4$, respectively.
 
Let us state the Ginibre and Jacobi ensemble as explicit examples of the new measure. In the Gaussian case, the integrals over $X'_j$ factorize and we find
\begin{align}\label{measureGaus1}
&d\nu_{I_{\rm L}}^{\rm (\gauss,L)}({X}^{(M)})\propto  \\
&d\mu(U_{\rm L})\prod_{j=1}^M\exp\left[-\tr \widetilde{X}_j\widetilde{X}_j^\dagger\right]{\det}^{\beta\nu_j/(2\gamma)}(\widetilde{X}_j\widetilde{X}_j^\dagger)d[\widetilde{X}_j], \nn
\end{align}
where the multi-index $I_{\rm L}=(\nu_1,\ldots,\nu_M)$ encodes the ordering of the exponents. Recall that $\nu_j=N_j-N_{\min}$. This measure was studied in Ref.~\cite{AkemannKieIps2013,Adhikari2013} for complex matrices ($\beta=2$) and in Ref.~\cite{Ipsen2013} for quaternion matrices ($\beta=4$).

For the Jacobi ensemble the integral~\eqref{newmeas3} is more involved. The Wishart matrix of the matrix $X_j$ has the form
\begin{equation}\label{Wishart}
 X_jX_j^\dagger=U_j\left(\left[\begin{array}{cc} \widetilde{X}_j\widetilde{X}_j^\dagger & 0 \\ 0 & 0 \end{array}\right]+X'_jX_j^{\prime\,\dagger}\right)U_j^\dagger
\end{equation}
The matrix $(\eins_{\gamma N_{\min}}-\widetilde{X}_j\widetilde{X}_j^\dagger)$ has to be positive definite, too. Therefore the transformation
\begin{equation}\label{trafo1}
 X'_j\to \left[\begin{array}{cc} (\eins_{\gamma N_{\min}}-\widetilde{X}_j\widetilde{X}_j^\dagger)^{1/2} & 0 \\ 0 & \eins_{\gamma (N_j-N_{\min})} \end{array}\right]X'_j
\end{equation}
is well-defined via the spectral decomposition theorem. Now, the integration over the rectangular matrices $X'_j\in{\rm gl}_\beta(N_j,\nu_{j-1})$ factorizes, and we end up with the new measure for the truncated unitary matrices
\begin{align}\label{measuretrunc1}
&d\nu_{\kappa,I_{\rm L}}^\text{(\jacobi,L)}({X}^{(M)})\propto d\mu(U_{\rm L})\\
&\times\prod_{j=1}^M \bigg[{\det}^{\kappa_j+\beta\nu_{j-1}/(2\gamma)}(\eins_{\gamma N_{\min}}-\widetilde{X}_j\widetilde{X}_j^\dagger) \nn\\
&\times{\det}^{\beta\nu_j/(2\gamma)}(\widetilde{X}_j\widetilde{X}_j^\dagger)\Theta(\eins_{\gamma N_{\min}}-\widetilde{X}_j\widetilde{X}_j^\dagger)d[\widetilde{X}_j]\bigg].\nn
\end{align}
Again, the subscripts denote $\kappa=(\kappa_1,\ldots,\kappa_M)$ and $I_{\rm L}=(\nu_1,\ldots,\nu_M)$, respectively. Notice that the new measure has a different exponent of the determinant $\det(\eins_{\gamma N_{\min}}-\widetilde{X}_j\widetilde{X}_j^\dagger)$ due to the integration over the rectangular matrices $X'_j$. This measure was studied in Ref.~\cite{AkemannKieBur2013} for complex matrices.

Let us return to general weights but now we consider the case $N_M=N_{\min}$. The matrix $U_{\rm L}$ is equal to unity. This time we start with $X_M$ and rotate it to the $\gamma N_{\min}\times \gamma N_{\min}$ matrix $\widetilde{X}_M$. Again we get a determinantal prefactor, which is now ${\det}^{\beta\nu_{M-1}/(2\gamma)}(\widetilde{X}_M\widetilde{X}_M^\dagger)$. We repeat the same procedure as described in the paragraphs above only starting from the left and ending at the right. We use a right block QR-decomposition (RQ-decomposition)
\begin{equation}\label{rightder}
 X_j=\left[\begin{array}{c} \begin{array}{cc} \widetilde{X}_j & 0 \end{array} \\ \hline X'_j\end{array}\right]U_j.
\end{equation}
The resulting measure is
\begin{equation}\label{measuregen2}
 d\nu_{\widetilde{P}_1^\text{(R)}\cdots\widetilde{P}^\text{(R)}_M}({X}^{(M)})= d\mu(U_{\rm R})\prod_{j=1}^M\widetilde{P}^\text{(R)}_j( \widetilde{X}_j)d[\widetilde{X}_j].
\end{equation}
with the individual weights given by
\begin{equation}\label{newmeas4}
 \widetilde{P}^\text{(R)}_j(\widetilde{X}_j )\propto{\det}^{\beta\nu_{j-1}/(2\gamma)}(\widetilde{X}_j\widetilde{X}_j^\dagger)\!\int\! d[X'_j] P_j\left(\left[\begin{array}{c} \begin{array}{cc} \widetilde{X}_j & 0 \end{array} \\ \hline X'_j\end{array}\right]\right).
\end{equation}
Here we integrate over the rectangular matrix $X'_j\in{\rm gl}_\beta(\nu_j,N_{j-1})$ and the superscript $(R)$ refers to the right block QR-decomposition~\eqref{rightder}. In particular, we have
\begin{align}\label{measureGaus2}
&  d\nu_{I_{\rm R}}^{\rm (\gauss,R)}({X}^{(M)})\propto  \\
&d\mu(U_{\rm R})\prod_{j=1}^M\exp\left[-\tr \widetilde{X}_j\widetilde{X}_j^\dagger\right]{\det}^{\beta\nu_{j-1}/(2\gamma)}(\widetilde{X}_j\widetilde{X}_j^\dagger)d[\widetilde{X}_j], \nn
\end{align}
for the Ginibre ensemble and 
\begin{align}\label{measuretrunc2}
&d\nu_{\kappa,I_R}^\text{(\jacobi,R)}({X}^{(M)})\propto d\mu(U_{\rm R}) \\
&\times\prod_{j=1}^M\biggl[{\det}^{\kappa_j+\beta\nu_{j}/(2\gamma)}(\eins_{\gamma N_{\min}}-\widetilde{X}_j\widetilde{X}_j^\dagger) \nn\\
&\times{\det}^{\beta\nu_{j-1}/(2\gamma)}(\widetilde{X}_j\widetilde{X}_j^\dagger)\Theta(\eins_{\gamma N_{\min}}-X_jX_j^\dagger)d[\widetilde{X}_j]\biggl].\nn
\end{align}
for the Jacobi ensemble, where the multi-index $I_{\rm R}=(\nu_0,\ldots,\nu_{M-1})$ encodes the order of the exponents. 

Note that the ``left'' and ``right'' measures~\eqref{newmeas2} and~\eqref{newmeas4} and therefore also Eqs.~\eqref{measuregen1} and~\eqref{measuregen2} differs by a replacement $\nu_{j}\leftrightarrow \nu_{j-1}$. How can we explain this discrepancy? And more importantly, does the case $N_M=N_0=N_{\min}$ yields a conflict, since both measures apply in this case? This problem can be easily resolved for Gaussian weights, see subsection~\ref{sec2.2}. In subsection~\ref{sec2.3}, we show a neat weak commutation relation between an induced weight of one random matrix reduced to a square matrix and the weight of a truncated unitary matrix (drawn from one of the three Jacobi ensembles). These two weak commutation relations are everything we need to understand the discrepancy between the measures~\eqref{measuregen1} and~\eqref{measuregen2}.

\subsection{A weak commutation relation of square random matrices of finite size}\label{sec2.2}

In this section we show that any ordering of the exponents of the determinantal prefactors in the Gaussian case yields the same statistics for the product matrix ${X}^{(M)}$. We also make  this statement stronger and show that any square random matrices distributed by probability densities invariant under the corresponding group commute inside the average. This weak commutation relation was already proven for matrices with infinite large matrix size \cite{Burda2013}, but is also exact at finite matrix size as we will show. Moreover it holds for all three Dyson classes. A weak commutation relation has also been discussed in the context of disordered wires with obstacles~\cite{BM:1994}.

Let $f$ be an integrable test function for the random matrix $Y=Y_2Y_1$ given as the product of two square random matrices $Y_{1,2}\in{\rm gl}_\beta(N_{\min},N_{\min})$ and distributed according to
\begin{equation}
 d\nu(Y)=p_1(Y_1)d[Y_1]p_2(Y_2)d[Y_2]\label{equivalence1}
\end{equation}
where the $p_{j}$ are probability densities invariant under $\U_\beta(N_{\min},N_{\min})$, i.e.
\begin{equation}
\int p_j(Y_j)d[Y_j]=1\quad\text{and}\quad
p_j(UY_jV)=p_j(Y_j).
\end{equation}
for all $U\otimes V\in\U_\beta(N_{\min},N_{\min})$. It follows that the integral over $f$ is invariant under $\U_\beta(N_{\min},N_{\min})$.
\begin{equation}\label{equivalence2}
 \int f(Y)d\nu(Y)=\int f(U_1YU_2)d\nu(Y),
\end{equation}
for all $U_1\otimes U_2\in\U_\beta(N_{\min},N_{\min})$. This is clear, since we can absorb $U_1$ and $U_2$ in the measures for $Y_2$ and $Y_1$, respectively. Integrating over $U_1$ and $U_2$ with respect to the normalized Haar measure on $\U_\beta(N_{\min},N_{\min})$, we define the function
\begin{equation}\label{deffunc}
 g(Y)= \int f(U_1YU_2)d\mu(U_1)d\mu(U_2).
\end{equation}
Indeed this auxiliary function only depends on the singular values of $Y$  due to the invariance under $\U_\beta(N_{\min},N_{\min})$. Additionally $Y$ and $Y^\dagger$ share the same singular values and lie in the same matrix space, namely ${\rm gl}_\beta(N_{\min},N_{\min})$. Hence, $g$ has the same functional dependence on $Y$ as on $Y^\dagger$. Thus the following relation holds
\begin{align}
 \int f(Y_1Y_2)d\nu(Y)&=\int g(Y)d\nu(Y) 
 =\int g(Y^\dagger)d\nu(Y) \nn \\
 &= \int f(Y_2^\dagger Y_1^\dagger)d\nu(Y). \label{equivalence3}
\end{align}
Finally we use the invariance of the measure $d\nu$ under the interchange $Y_j\leftrightarrow Y_j^\dagger$, which yields the main result of this section,
\begin{equation}\label{comrelsquare}
 \int f(Y_1Y_2)d\nu(Y)= \int f(Y_2 Y_1)d\nu(Y).
\end{equation}
Thus the two random matrices commute in a weak sense.

For example, let the measures be deformed Gaussians,
\begin{equation}
 P_j(Y_j)=\frac1{Z_j} {\det}^{\nu_j} (Y_jY_j^\dagger) \exp[-\tr Y_jY_j^\dagger],
\end{equation}
where $Z_j$ is a normalization constant.
Then the measure of $Y=Y_1Y_2$ is invariant under the interchange of the two exponents $\nu_1$ and $\nu_2$ in the measure of $Y_1$ and $Y_2$. Applying this knowledge to the measure~\eqref{measureGaus1} we can show that any two neighbouring matrices can be interchanged. These interchanges are the generators of the permutation group $\mathbb{S}(M)$ of $M$ elements. It follows that 
\begin{equation}\label{equivalence5}
 \int f({X}^{(M)})d\nu_{I_{\rm L}}^{\rm (\gauss)}({X}^{(M)})=
 \int f({X}^{(M)})d\nu_{\omega(I_{\rm L})}^{\rm (\gauss)}({X}^{(M)}),
\end{equation}
for all $\omega\in\mathbb{S}(M)$ and any integrable test-function $f$ of the random matrix ${X}^{(M)}$. Therefore the ordering of the multi-index $I_{\rm L}$ is irrelevant. The equivalence relation will be indeed reflected in the discussion of the eigenvalue statistics in Sec.~\ref{sec3.2}.

A na\"ive generalization of the weak commutation relation~\eqref{comrelsquare} to any two rectangular random matrices does not work since in general the matrix dimensions does not close, meaning that two matrices might be multipliable like $Y_1Y_2$ but not like $Y_2Y_1$.

\subsection{A weak commutation relation of square random matrices with an induced measure}\label{sec2.3}

Let us consider a second weak commutation relation since the former commutation relation only solves the problem of an ambiguity of the resulting weight of ${X}^{(M)}$ when the integrals over $X_j^\prime$ factorize from the rest, as for Gaussian weights. What happens with other random matrix ensembles? To answer this question we consider the rectangular random matrix $Y\in{\rm gl}_\beta(N_1,N_0)$ distributed by the weight $P$. Furthermore, we assume that $f$ is an arbitrary integrable function on the set ${\rm gl}_\beta(N_{\min},N_{\min})$ with $N_0,N_1\geq N_{\min}$.

We consider the integral
\begin{equation}\label{2.3.1}
 I[f]=\int f\left(X_{\rm L}Y_{\rm L}\right)P_{\rm L}(Y_{\rm L})d[Y_{\rm L}] d\nu_{\kappa_1}^{\rm(\jacobi)}(X_{\rm L}),
\end{equation}
where $\kappa_1=\beta(N_1-2N_{\min}+1-2/\beta)/2\gamma$, cf. Eq.~\eqref{Jacobimeas}. The truncated unitary matrix $X_{\rm L}\in{\rm gl}_\beta(N_{\min},N_{\min})$ is distributed according to the Jacobi measure
\begin{multline}\label{2.3.2}
d\nu_{\kappa_1}^{\rm(\jacobi)}(X_{\rm L})\propto \Theta(\eins_{\gamma N_{\min}}-X_{\rm L}X_{\rm L}^\dagger)\\
\times{\det}^{\kappa_1}(\eins_{\gamma N_{\min}}-X_{\rm L}X_{\rm L}^\dagger)d[X_{\rm L}],
\end{multline}
cf. Eq.~\eqref{Jacobimeas}. The measure $P_{\rm L}$ is the induced measure
\begin{equation}\label{2.3.3}
 P_{\rm L}(Y_{\rm L})\propto{\det}^{\beta\nu_1/(2\gamma)}(Y_{\rm L}Y_{\rm L}^\dagger)\!\int\! d[Y'] P\left(\left[\begin{array}{c|c} \begin{array}{c} Y_{\rm L} \\ 0 \end{array} & Y'\end{array}\right]\right),
\end{equation}
for the sub-block $Y_{\rm L}\in{\rm gl}_\beta(N_{\min},N_{\min})$, where we integrate over the rectangular matrix $Y'\in{\rm gl}_\beta(N_1,\nu_0)$, cf. Eq.~\eqref{newmeas3}.

The random matrix $X_{\rm L}$ can be written as a product of three matrices. Two of them are projections which restrict the group element $U_{\rm L}\in\mathbb{G}_\beta(N_1,N_{\min})$ to its first $\gamma N_{\min}$ rows and columns,
\begin{equation}\label{2.3.4}
 X_{\rm L}=\left[\begin{array}{cc} \eins_{\gamma N_{\min}} & 0\end{array}\right]U_{\rm L}\left[\begin{array}{c} \eins_{\gamma N_{\min}} \\ 0\end{array}\right],
\end{equation}
 where $U_{\rm L}$ is weighted with respect to the Haar measure on $\mathbb{G}_\beta(N_1,N_{\min})$. Employing the inverse decomposition, see Eq.~\eqref{leftder3},
\begin{equation}\label{2.3.5}
 Y=U_{\rm L}\left[\begin{array}{c|c} \begin{array}{c} Y_{\rm L} \\ 0 \end{array} & Y'\end{array}\right].
\end{equation}
We can rewrite the integral~\eqref{2.3.1} in terms of an integral over $Y$ distributed according to the density $P$,
\begin{align}\label{2.3.6}
 I[f]&=\int f\left(\left[\begin{array}{cc} \eins_{\gamma N_{\min}} & 0\end{array}\right]U_{\rm L}\left[\begin{array}{c} Y_{\rm L} \\ 0\end{array}\right]\right)P_{\rm L}(Y_{\rm L})d[Y_{\rm L}] d\mu(U_{\rm L})\nn\\
 &=\int f\left(\left[\begin{array}{cc} \eins_{\gamma N_{\min}} & 0\end{array}\right]Y\left[\begin{array}{c} \eins_{\gamma N_{\min}} \\ 0\end{array}\right]\right)P(Y)d[Y] .
\end{align}
This procedure can be inverted only with the difference that we do it to the right, i.e. we consider the decomposition 
\begin{equation}\label{2.3.7}
 Y=\left[\begin{array}{c} \begin{array}{cc} Y_{\rm R} & 0 \end{array} \\ \hline Y''\end{array}\right]U_{\rm R}.
\end{equation}
with $Y_{\rm R}\in{\rm gl}_\beta(N_{\min},N_{\min})$ and the truncated matrix
\begin{equation}\label{2.3.8}
 X_{\rm R}=\left[\begin{array}{cc} \eins_{\gamma N_{\min}} & 0\end{array}\right]U_{\rm R}\left[\begin{array}{c} \eins_{\gamma N_{\min}} \\ 0\end{array}\right]\in{\rm gl}_\beta(N_{\min},N_{\min}),
\end{equation}
which induces the measures
\begin{multline}\label{2.3.9}
d\nu_{\kappa_0}^{\rm(\jacobi)}(X_{\rm R})\propto \Theta(\eins_{\gamma N_{\min}}-X_{\rm R}X_{\rm R}^\dagger) \\
\times{\det}^{\kappa_0}(\eins_{\gamma N_{\min}}-X_{\rm R}X_{\rm R}^\dagger)d[X_{\rm R}]
\end{multline}
with $\kappa_0=\beta(N_0-2N_{\min}+1-2/\beta)/2\gamma$ and
\begin{equation}\label{2.3.10}
 P_{\rm R}(Y_{\rm R})\propto{\det}^{\beta\nu_0/(2\gamma)}(Y_{\rm R}Y_{\rm R}^\dagger)\!\int\! d[Y''] P\left(\left[\begin{array}{c} \begin{array}{cc} Y_{\rm R} & 0 \end{array} \\ \hline Y''\end{array}\right]\right),
\end{equation}
where we integrate over the rectangular matrix $Y''\in{\rm gl}_\beta(\nu_1,N_0)$. Thus we find the following identity
\begin{align}\label{2.3.11}
  I[f]&=\int f\left(X_{\rm L}Y_{\rm L}\right)P_{\rm L}(Y_{\rm L})d[Y_{\rm L}] d\nu_{\kappa_1}^{\rm(\jacobi)}(X_{\rm L})\nonumber\\
 &=\int f\left(Y_{\rm R}X_{\rm R}\right)P_{\rm R}(Y_{\rm R})d[Y_{\rm R}] d\nu_{\kappa_0}^{\rm(\jacobi)}(X_{\rm R})
\end{align}
which is the main result of this section. This identity can be considered as some kind of weak commutation relation for induced square matrices with truncated unitary random matrices. It is the missing link between the effective weights~\eqref{measuregen1} and \eqref{measuregen2}.

\subsection{The general case}\label{sec2.4}

Let us consider the general product matrix, $X^{(M)}=X_M\cdots X_1$ where $N_J=N_{\min}$ for some $J\in\{0\ldots M\}$. It is irrelevant whether $J$ is unique or not, due to the weak commutation relations discussed in subsections~\ref{sec2.2} and \ref{sec2.3}. If there is more than one $J$ such that $N_J=N_{\min}$, then one can take any of these. In particular the identity~\eqref{2.3.11} allows us to transform the resulting measure for $X^{(M)}$ to equivalent weights.

We define the sub-product matrices
\begin{align}
 X^{(M, {\rm L})}&=X_MX_{M-1}\cdots X_{J+2}X_{J+1}\in{\rm gl}_\beta(N_{M},N_{\min}), \nn\\
 X^{(M, {\rm R})}&=X_JX_{J-1}\cdots X_2X_1\in{\rm gl}_\beta(N_{\min},N_0),
\end{align}
such that $X^{(M)}=X^{(M, {\rm L})} X^{(M, {\rm R})}$. We apply the procedure for deriving the measure~\eqref{measuregen1} on the matrix $X^{(M, {\rm L})}$ and the procedure for the measure~\eqref{measuregen2} on the matrix $X^{(M, {\rm R})}$. Then we obtain the measure and the main result of this section
\begin{equation}\label{measureresult}
 d\nu_{\widetilde{P}_1\ldots \widetilde{P}_M}({X}^{(M)})= d\mu(U_{\rm L})d\mu(U_{\rm R})\prod_{j=1}^M\widetilde{P}_j( \widetilde{X}_j)d[\widetilde{X}_j],
\end{equation}
where $\widetilde{P}_j$ and $\widetilde{X}_j$ are given by Eqs.~\eqref{newmeas3} and \eqref{leftder3} for $j>J$ and by Eqs.~\eqref{newmeas4} and \eqref{rightder} for $j\leq J$.

For the Gaussian case we obtain
\begin{align}\label{measureGaus3}
 & d\nu_{I}^{\rm (\gauss)}({X}^{(M)})\\
 & \propto d\mu(U_{\rm L})\prod_{j=J+1}^M\exp\left[-\tr \widetilde{X}_j\widetilde{X}_j^\dagger\right]{\det}^{\beta\nu_{j}/(2\gamma)}(\widetilde{X}_j\widetilde{X}_j^\dagger)d[\widetilde{X}_j] \nn\\
 & \times d\mu(U_{\rm R})\prod_{j=1}^J\exp\left[-\tr \widetilde{X}_j\widetilde{X}_j^\dagger\right]{\det}^{\beta\nu_{j-1}/(2\gamma)}(\widetilde{X}_j\widetilde{X}_j^\dagger)d[\widetilde{X}_j]\nonumber
\end{align}
with the multi index $I=(\nu_0,\ldots, \nu_{J-1},\nu_{J+1},\ldots, \nu_M)$. Indeed we can again apply the weak commutation relation~\eqref{equivalence5} to the product matrix ${X}^{(M)}$ which tells us that we can take any permutation of the multi-index $I$ and, hence, of the exponents of the determinantal prefactors.

The effective measure for a product of matrices drawn from Jacobi ensembles yields  a result similar to Eq.~\eqref{measureGaus3}. We have only to plug  in the measures~\eqref{measuretrunc1} for $X^{(M, {\rm L})}$ and \eqref{measuretrunc2} for $X^{(M, {\rm R})}$ instead of the deformed Gaussians.

We emphasize that the effective measure~\eqref{measureresult} does not only apply for the discussion of the eigenvalue or singular value  statistics. One can also apply this reduction to a product of square matrices to correlations of the eigenvectors as well as to cross-correlations of eigenvectors and eigenvalues. Moreover the permutation invariance due to the commutation relation~\eqref{comrelsquare} and the choice of $J$, if more than one $N_j$ is equal to the minimal dimension $N_{\min}$, has to be obviously reflected in the statistics of the eigenvalues and singular values (c.f. Ref.~\cite{AkemannKieIps2013}) of $\widetilde{X}^{(M)}$ and, thus, of $X^{(M)}$. This has crucial physical implications. Consider the chaotic quantum system from Sec.~\ref{sec1}, the permutation symmetry tells us that the order of the potential wells is completely irrelevant, if we consider only the statistics of the product matrix $X^{(M)}$ involved in these problems. Note that the cross-correlations between particular 
positions of the chain are 
still affected by the order of the potential wells.

\section{Eigenvalue statistics for products of matrices}\label{sec3}

In this section we will discuss the eigenvalues of $X^{(M)}=X_M\cdots X_1$, which are of general interest in a broad spectrum of applications. Due to the discussion of Sec.~\ref{sec2} we can restrict ourselves to the case of square matrices, $N_0=\ldots=N_M=N_{\min}=N$, without loss of generality. The matrices $X_j$ are square matrices with different weights, which can be chosen to imitate a product of rectangular matrices. We are interested in the statistical properties of the eigenvalues of the product matrix $X^{(M)}\in{\rm gl}_\beta(N,N)$, see Eq.~\eqref{defproduct}. Thus we are looking for the zeros of the characteristic polynomial
\begin{equation}\label{charaeq}
 \det(X^{(M)}-z\eins_{\gamma N})=\det(X_M\cdots X_2 X_1-z\eins_{\gamma N})=0.
\end{equation}

In subsection~\ref{sec3.1} we perform an eigenvalue decomposition for arbitrary weights of the matrices $X_j$. We specify this decomposition for the Gaussian case in subsection~\ref{sec3.2} and for the Jacobi ensemble in subsection~\ref{sec3.3} and calculate an explicit expression for the joint probability density function for both ensembles.

\subsection{The eigenvalue decomposition}\label{sec3.1}

We pursue the idea of the Ginibre ensembles for the Dyson indices $\beta=2,4$ employed by the authors of Refs.~\cite{AkemannBur2012,Ipsen2013} and perform a generalized Schur decomposition. Let
\begin{equation}\label{2.1}
B=X\widehat{T}
\end{equation}
and $X=\diag(X_1,\ldots,X_M)$ a block-diagonal matrix and $\widehat{T}$ the constant matrix as in Eq.~\eqref{translation}. Then $B^M$ has the same eigenvalues as $X^{(M)}$ only that they are $M$ times degenerate.

\subsubsection{$\beta=2$}\label{sec3.1.1}

We first consider the simplest case $\beta=2$. In the first step we perform a simultaneous decomposition of $B$ in a diagonal matrix $Z=\diag(Z_1,\ldots,Z_M)$ with complex eigenvalues $Z_j=\diag(z_1^{(j)},\ldots,z_N^{(j)})$, an upper triangular matrix $\Delta=\diag(\Delta_1,\ldots,\Delta_M)$ and a unitary matrix  $U=\diag(U_1,\ldots,U_M)$. We have
\begin{equation}\label{2.2-bet2}
 B=X\widehat{T}=U(Z+\Delta)\widehat{T}U^\dagger.
\end{equation}
The differential of $B$ is given by
\begin{equation}\label{2.3-bet2}
 dB=dX\widehat{T}=U\left((dZ+d\Delta)\widehat{T}+[dA,(Z+\Delta)\widehat{T}]_-\right)U^\dagger,
\end{equation}
where $dA=U^\dagger dU$. The differentials of $Z$ and $\Delta$ completely factorize from the rest. Only the $Z$-dependent part in the commutator, $[\,\cdot\,,\,\cdot\,]_-$, contributes to the Jacobian. The upper triangular matrix $\Delta$ incorporates a recursive shift of $dA$ which results in a upper triangular part of the Jacobian, as well. The variable $dA_{ab}^{(j)}$ denotes a matrix element of the $j$-th matrix $dA_j=U_j^\dagger dU_j$ which is a complex variable. Moreover $dA_{aa}^{(j)}=0$ since these degrees of freedom are incorporated in $Z$, hence
\begin{equation}\label{2.4-bet2}
 \{[dA,Z\widehat{T}]_-\widehat{T}^\dagger\}_{ab}^{(j)}=dA_{ab}^{(j)}z_b^{(j)}-z_a^{(j)}dA_{ab}^{(j+1)}.
\end{equation}
The Jacobian resulting from this transformation is a determinant with a diagonal part corresponding to $dZ$ and $d\Delta$ and a part proportional to $\widehat{T}$ resulting from $dA$. Then we arrive at
\begin{equation}\label{2.5-bet2}
 \prod_{j=1}^Md[X_j]\propto\left|\Delta_N\left(Z_M\cdots Z_1\right)\right|^2
\prod_{j=1}^Md\mu(U_j)d[Z_j]d[\Delta_j].
\end{equation}
The differential for $Z_j$ is $d[Z_j]=\prod_{a=1}^Nd\RE\,z_a^{(j)}d\IM\,z_a^{(j)}$, while $\Delta_N(Z)$ denotes the Vandermonde determinant.

Let us return to the full measure where each matrix $X_j$ is distributed via the probability density $P_j$. Due to the factorization of the differentials $d[\Delta_j]$ we define the reduced weights
\begin{equation}\label{neww-bet2}
 \widehat{P}_j(Z_j)\equiv\int P_j(Z_j+\Delta^{(j)})d[\Delta^{(j)}].
\end{equation}
The joint probability density of the eigenvalues of the product matrix $X^{(M)}$ reads in these new weights
\begin{multline}\label{jpdfgen-bet2}
p^{ (\beta=2)}(Z^{(M)})\propto |\Delta_N(Z^{(M)})|^2 \\
\times\prod_{j=1}^M\int d[Z_j] \widehat{P}_j(Z_j)\prod_{a=1}^{N}\delta^{(2)}(z_a-z_{a}^{(M)}\cdots z_{a}^{(1)})
\end{multline}
where we use the Dirac $\delta$-function for complex variables, $\delta^{(2)}(z)=\delta(\RE\, z)\delta(\IM\, z)$, and define the diagonal product matrix $Z^{(M)}=Z_1\cdots Z_M$. Expression~\eqref{jpdfgen-bet2} is the farthest one can calculate for an arbitrary weight. If one wants to have a more concrete result one has to specify the measures $\widehat{P}_j$. We will do this for the Gaussian measure in subsection~\ref{sec3.2} and for the Jacobi measure in subsection~\ref{sec3.3} and recover the results derived in Refs.~\cite{AkemannBur2012,Adhikari2013,AkemannKieBur2013}.

\subsubsection{$\beta=4$}\label{sec3.1.2}

The next case we consider is $\beta=4$. In this case we can again decompose $B$ in an upper triangular matrix $\Delta$, a unitary symplectic matrix $U\in\USp(2N)$ and a $2N\times2N$ matrix $\widehat{Z}=\diag(Z,Z^*)$ where $Z$ is the same complex, diagonal $N\times N$ matrix as in the case $\beta=2$. We replace $Z\to\widehat{Z}$ in Eqs.~(\ref{2.3-bet2}\,-\,\ref{neww-bet2}). Moreover the complex matrix elements in the case $\beta=2$, $\{\Delta_j\}_{ab}$ and $dA_{ab}^{(j)}$ with $a\neq b$,  are now $2\times 2$ quaternion matrix blocks with four real independent  elements. Each of the diagonal $2\times2$ blocks $dA_{aa}^{(j)}$ only contain one off-diagonal complex variable. Hence the analogue to Eq.~\eqref{2.4-bet2} is
\begin{multline}\label{2.4-bet4}
 \{[dA,Z\widehat{T}]_-\widehat{T}^\dagger\}_{ab}^{(j)}=\\
dA_{ab}^{(j)} \begin{pmatrix}z_b^{(j)} & 0 \\ 0 & z_b^{(j)*}\end{pmatrix}-
\begin{pmatrix}z_a^{(j)} & 0 \\ 0 & z_a^{(j)*}\end{pmatrix}dA_{ab}^{(j+1)}
\end{multline}
The computation of the Jacobian works exactly the same as in the case $\beta=2$ and we find the joint probability density of the eigenvalues
\begin{multline}\label{jpdfgen-bet4}
 p^{ (\beta=4)}(Z^{(M)})\propto\Delta_{2N}(Z^{(M)},{Z^{(M)}}^*)\prod_{a=1}^{ N}(z_a-z_a^*)\\
 \times\prod_{j=1}^M\int d[Z_j] \widehat{P}_j(Z_j) \prod_{b=1}^{ N}\delta^{(2)}(z_b-z_{b}^{(M)}\cdots z_{b}^{(1)})
\end{multline}
with
\begin{equation}\label{neww-bet4}
 \widehat{P}_j(Z_j)\propto\int P_j(\diag(Z_j,Z_j^*)+\Delta^{(j)})d[\Delta^{(j)}].
\end{equation}
Again one needs specific weights $P_j$ to calculate further. For the Gaussian case the resulting measure was studied by one of the authors in Ref.~\cite{Ipsen2013}.

\subsubsection{$\beta=1$}\label{sec3.1.3}

Finally let us consider the case $\beta=1$. We have to distinguish between odd and even matrix dimensions. For this reason we introduce the notation $N=2\widetilde{N}+\chi$ with $\chi=0$ or $\chi=1$. Unlike the complex Schur decomposition, the real Schur decomposition will not generally trace $B$ back to a triangular form; instead similarity transformations with orthogonal matrices $U\in\Ort(2\widetilde{N}+\chi)$ bring $B$ to a block diagonal matrix $\widehat{Z}=\diag(\widehat{Z}_1^{(1)},\widehat{Z}_2^{(1)},\ldots,\widehat{Z}_{\widetilde{N}+\chi}^{(1)},\widehat{Z}_1^{(2)},\ldots,\widehat{Z}_{\widetilde{N}+\chi}^{(M)})$ and an upper block triangular matrix $\Delta$~\cite{Loan}.  The blocks $\widehat{Z}_1^{(1)},\ldots,\widehat{Z}_{\widetilde{N}}^{(M)}$ are $2\times2$ real matrices and $\widehat{Z}_{\widetilde{N}+1}^{(j)}$ (only for $\chi=1$) is a real number. Thus the matrix elements of $\Delta$ and $dA_{ab}^{(j)}$, $a\neq b$ and $a,b\neq\widetilde{N}+1$, in the case $\beta=2$ are again replaced by $2\times 2$ 
matrix blocks 
with four real independent variables. 
The block diagonal elements $dA_{aa}^{(j)}$ are zero. For $\chi=1$ we have two additional real variables arranged in a two dimensional vector $dA_{a, \widetilde{N}+1}^{(j)}$ for each $a=1\ldots\widetilde{N}$ and $j=1\ldots M$. 
In the case of an even dimensional matrix ($\chi=0$) the differentials are
\begin{equation}
 \{[dA,Z\widehat{T}]_-\widehat{T}^\dagger\}_{ab}^{(j)}=dA_{ab}^{(j)}\widehat{Z}_b^{(j)}- \widehat{Z}_a^{(j)}dA_{ab}^{(j+1)}
\label{differential-real}
\end{equation}
as in Eq.~\eqref{2.4-bet2} but with $2\times 2$ real matrices. For odd dimensional matrices ($\chi=1$) one needs to treat the case where $a$ or $b$ are equal to $\widetilde N+1$ separately, then the $2\times 2$ real matrices $dA_{ab}^{(j)}$ in Eq.~\eqref{differential-real} have to be replaced by a chain of two dimensional real vectors, $dA_{a,\widetilde N+1}^{(j)}$ or $dA_{\widetilde N+1,b}^{(j)}$, which is in the spirit of Ref.~\cite{Sommers2008}.
\begin{widetext}
Let $\widetilde{Z}_a^{(M)}=\widehat{Z}_a^{(M)}\cdots \widehat{Z}_a^{(1)}$ and $\widetilde{Z}^{(M)}=\widehat{Z}^{(M)}\cdots \widehat{Z}^{(1)}$. Then the resulting joint probability density of the matrix blocks $\widetilde{Z}_a^{(M)}$ is
\begin{equation}\label{jpdfgen-bet1.ev}
 \widehat{p}^{ (\beta=1)}(\widetilde{Z}^{(M)})\propto \prod_{1\leq a<b\leq\widetilde{N}}\left|\det\left[\widetilde{Z}_a^{(M)}\otimes\eins_{2}-\eins_2\otimes \widetilde{Z}_b^{(M)}\right]\right|\prod_{j=1}^M\int d[\widehat{Z}^{(j)}] \widehat{P}_j(\widehat{Z}^{(j)})
\prod_{a=1}^{\widetilde{N}}\delta^{(4)}\left(\widetilde{Z}_a^{(M)}-\widehat{Z}_a^{(M)}\cdots \widehat{Z}_a^{(1)}\right)\nonumber
\end{equation}
for an even dimension $N$ and 
\begin{multline}
 \widehat{p}^{ (\beta=1)}(\widetilde{Z}^{(M)})\propto \prod_{1\leq a<b\leq\widetilde{N}}\left|\det\left[\widetilde{Z}_a^{(M)}\otimes\eins_{2}-\eins_2\otimes \widetilde{Z}_b^{(M)}\right]\right|\prod_{j=1}^{\widetilde{N}}\left|\det\left[\widetilde{Z}_j^{(M)}-\widetilde{Z}_{\widetilde{N}+1}^{(M)}\eins_2\right]\right|\label{jpdfgen-bet1.odd}\\
\times\prod_{j=1}^M\int d[\widehat{Z}^{(j)}] \widehat{P}_j(\widehat{Z}^{(j)}) \delta^{(2)}\left(\widetilde{Z}_{\widetilde{N}+1}-\widehat{Z}_{\widetilde{N}+1}^{(M)}\cdots \widehat{Z}_{\widetilde{N}+1}^{(1)} \right)
\prod_{a=1}^{\widetilde{N}}\delta^{(4)}\left(Z_a^{(M)}-\widehat{Z}_a^{(M)}\cdots \widehat{Z}_a^{(1)}\right)
\end{multline}
for an odd one. The first product of determinants incorporates the differences of pairs of $2\times2$ matrices.
Therefore those determinants are over $4\times4$ matrices and are reminiscent of a Vandermonde determinant.
The Dirac $\delta$-function over a $2\times 2$ real matrix is the product of the Dirac $\delta$-functions of all four real independent variables. The reduced probability densities are defined as always,
\begin{equation}\label{neww-bet1}
 \widehat{P}_j(\widehat{Z}^{(j)})\equiv\int P_j(\widehat{Z}^{(j)}+\Delta^{(j)})d[\Delta^{(j)}].
\end{equation}
The only difference of this definition to Eqs.~\eqref{neww-bet2} and \eqref{neww-bet4} is that we remain with a distribution for  block diagonal matrices, $\widehat{Z}^{(j)}$, instead with diagonal ones.

The eigenvalues of the $2\times 2$ real matrices $\widetilde{Z}_a^{(M)}$ are either a complex conjugate pair or two independent real eigenvalues. Since the probability densities $\widehat{P}_j$ are invariant under left and right multiplication of $\Ort(N)$ we can replace the argument $\widetilde{Z}^{(M)}$ of the recursive integral (for simplicity only shown for even dimension, $\chi=0$)
\begin{equation}\label{recint.1}
I(\widetilde{Z}^{(M)})= \prod_{j=1}^M\int d[\widehat{Z}^{(j)}] \widehat{P}_j(\widehat{Z}^{(j)})\prod_{a=1}^{\widetilde{N}}\delta^{(4)}(\widetilde{Z}_a^{(M)}-\widehat{Z}_a^{(M)}\cdots \widehat{Z}_a^{(1)})
\end{equation}
by the positive definite matrix $\sqrt{\widetilde{Z}^{(M)}\widetilde{Z}^{(M)\,\dagger}}$. This matrix is a block diagonal matrix which can be readily expressed by a singular value decomposition $\widetilde{Z}^{(M)}=U_{\rm L}\Lambda^{(M)} U_{\rm R}$ with $U_{\rm L},U_{\rm R}\in\Ort(2)^{\widetilde{N}}$ and $\Lambda$ a positive diagonal matrix. The idea is to calculate an integral representation of the joint probability distribution of the eigenvalues of $\widetilde{Z}^{(M)}$ in terms of the singular values in $\Lambda^{(M)}$. For this purpose it is quite helpful that we could reduce the whole problem to a $2\times2$ matrix problem.

In appendix~\ref{app1} we derived a general relation between the eigenvalues and the singular values of a $2\times2$ real matrix. Employing the result of the calculations~(\ref{jointp2}\,-\,\ref{jointp4}) we find the joint probability density of the eigenvalues $Z^{(M)}=\diag(z_1,\ldots,z_N)$ of $X^{(M)}$,
\begin{multline}\label{jpdfgen-bet1.ev.2}
 p^{ (1)}(Z^{(M)})\propto |\Delta_{2\widetilde{N}}(Z^{(M)})|\prod_{j=1}^M\int d[\widehat{Z}^{(j)}] \widehat{P}_j(\widehat{Z}^{(j)})
\bigg[\prod_{a=1}^{\widetilde{N}}\left(\delta(\IM\,z_{2a-1})\delta(\IM\,z_{2a})+2\delta^{(2)}(z_{2a-1}-z_{2a}^*)\right)\\
\times\int_{\alpha}^\infty d\alpha_a \delta^{(4)}(\Lambda(z_{2a-1},z_{2a},\alpha_a)-\widehat{Z}_a^{(M)}\cdots \widehat{Z}_a^{(1)})\bigg]
\end{multline}
for even dimension and 
\begin{multline}\label{jpdfgen-bet1.odd.2}
 p^{ (1)}(Z^{(M)})\propto |\Delta_{2\widetilde{N}+1}(Z^{(M)})|\prod_{j=1}^M\int d[\widehat{Z}^{(j)}] \widehat{P}_j(\widehat{Z}^{(j)})\delta\left(\IM\,z_{2\widetilde{N}+1}\right) \delta\left(\RE\,z_{2\widetilde{N}+1}-\widehat{Z}_{2\widetilde{N}+1}^{(M)}\cdots \widehat{Z}_{2\widetilde{N}+1}^{(1)}\right)\\
\times \bigg[\prod_{a=1}^{\widetilde{N}}\left(\delta(\IM\,z_{2a-1})\delta(\IM\,z_{2a})+2\delta^{(2)}(z_{2a-1}-z_{2a}^*)\right)
\int_{\alpha}^\infty  d\alpha_a \delta^{(4)}(\Lambda(z_{2a-1},z_{2a},\alpha_a)-\widehat{Z}_a^{(M)}\cdots \widehat{Z}_a^{(1)})\bigg]
\end{multline}
for odd dimension, where $\alpha=|\IM(z_{2a-1}-z_{2a})|/2$. In both cases we employed the functional dependence of the singular values on the eigenvalues, i.e.
\begin{align}
\Lambda(z_{2a-1},z_{2a},\alpha_a)=\begin{bmatrix} \lambda_+(z_{2a-1},z_{2a},\alpha_a) & 0 \\ 0 & \lambda_-(z_{2a-1},z_{2a},\alpha_a) \end{bmatrix}
=\sqrt{\alpha_a^2+\frac{(z_{2a-1}+z_{2a})^2}{4}}\eins_2+\sqrt{\alpha_a^2+\frac{(z_{2a-1}-z_{2a})^2}{4}}\sigma_3
\label{singfunc}
\end{align}
cf. Eq.~\eqref{app.sing}. The $2\times2$ matrix $\sigma_3$ is the third Pauli matrix. The integrals over $\alpha_a$ are reminiscent to the integrals found in the real Ginibre ensemble (equal to the case $M=1$) generating the error function \cite{Sommers2008} and in the real chiral Ginibre ensemble (equal to the case $M=2$) yielding an integral over a Bessel function \cite{AkemannPhi2010}. Also the prefactor consisting of the Dirac $\delta$-function is the same in both cases and is a universal factor reflecting the nature of the eigenvalues of arbitrary real matrices.

An important remark is in order. Assuming  one of the pairs of eigenvalues is real, say $(z_{2N-1},z_{2N})$, one can also approach an eigenvalue decomposition of the product of $2\times 2$ matrices by a generalized Schur decomposition. Thus the following integral over $\widehat{Z}_N^{(j)}$ is equivalent
\begin{multline}\label{intidentity-real}
\delta(\IM\,z_{2N-1})\delta(\IM\,z_{2N})\prod_{j=1}^M\int d[\widehat{Z}_N^{(j)}] \widehat{P}_j\left(\left[\begin{array}{ccc} \widehat{Z}_1^{(j)} & & 0 \\ & \ddots & \\ 0 & & \widehat{Z}_N^{(j)} \end{array}\right]\right)
\int_{0}^\infty d\alpha_N \delta^{(4)}\left(\Lambda(z_{2N-1},z_{2N},\alpha_N)-\widehat{Z}_N^{(M)}\cdots \widehat{Z}_N^{(1)}\right)\\
\propto\delta(\IM\,z_{2N-1})\delta(\IM\,z_{2N})\prod_{j=1}^M\int dx_1^{(j)} dx_2^{(j)} dx_3^{(j)} \widehat{P}_j\left(\left[\begin{array}{ccc|c} \widehat{Z}_1^{(j)} & & 0 &  \\ & \ddots & & 0 \\ 0 & & \widehat{Z}_{N-1}^{(j)}  \\ \hline  & 0 & & \begin{array}{cc} x_1^{(j)} & x_2^{(j)} \\ 0 & x_3^{(j)} \end{array} \end{array}\right]\right)\\
\times\delta\left(\RE\,z_{2N-1}-x_1^{(M)}\cdots x_1^{(1)}\right)\delta\left(\RE\,z_{2N}-x_3^{(M)}\cdots x_3^{(1)}\right).
\end{multline}
The integral on the right hand side was used quite recently in Ref.~\cite{Forrester2013} to calculate the probability of a fixed number of real eigenvalues for a product of Ginibre matrices. Notice that the integral identity~\eqref{intidentity-real} is not at all trivial and we know only that it has to be in general true since both approaches are legitimized.

Also in the case of real matrices we need a specific measure to calculate any further. This is exactly what we do in the next two subsections and restrict our discussion to the Ginibre and Jacobi ensemble. We emphasize that the discussion so far have been for completely arbitrary probability weights and can be applied to a broad class of ensembles.

\end{widetext}

\subsection{Products of Ginibre matrices}\label{sec3.2}

As discussed in Sec.~\ref{sec2}, a product of rectangular Ginibre matrices, $X^{(M)}=X_M\cdots X_1$, is closely related to a product of square matrices, see Eq.~\eqref{measureGaus1}. Applying a Schur decomposition, the rotations are trivially integrated out and they contribute only to the normalization. Likewise, the triangular matrices from the Schur decomposition completely drops out in the determinantal prefactor and factorizes in the Gaussian part, such that also these integrals result in a constant.

Let us again restrict ourselves to complex matrices ($\beta=2$) first. Starting from Eqs.~\eqref{jpdfgen-bet2} and \eqref{measureGaus1}, the joint probability density is
\begin{equation}\label{joint-2}
  p_{{ \nu}}^{( {\rm \gauss}, \beta=2)}(Z^{(M)})\propto |\Delta_{N_{\min}}(Z^{(M)})|^2\prod_{a=1}^{N_{\min}}g^{({\rm \gauss}, M)}_{{ \nu}}(z_a)
\end{equation}
with the one-point weight
\begin{multline}\label{weight-2a}
g^{( {\rm \gauss}, M)}_{{ \nu}}(z)= \prod_{j=1}^{M}\int_{\mathbb{C}} d^2z^{(j)} |z^{(j)}|^{2\nu_j}e^{-|z^{(j)}|^2}\\
\times\delta^{(2)}(z-z^{(M)}\cdots z^{(1)}).
\end{multline}
We employ the abbreviation ${ \nu}=(\nu_1,\nu_2,\ldots,\nu_M)$. The integral is equal to a representation of a  Meijer $G$-function \cite{Gradshteynbook}
\begin{align}\label{weight-2b}
 g^{ ({\rm \gauss}, M)}_{{\nu}}(z)&= \MeijerG{0}{M}{M}{0}{-}{\nu_1,\ldots,\nu_M}{\abs z^2}\\
 &=\int_{\mathcal{C}}\frac{du}{2\pi\imath}|z|^{2u}\prod_{j=1}^M\Gamma(\nu_j-u)\nonumber
\end{align}
The second line is a quite useful integral representation of the Meijer $G$-function, where the contour $\mathcal{C}$ runs around the poles of the gamma functions. Recall that the invariance under permutations of the indices, $\nu_j=N_j-N_{\min}$ reflects the weak commutation relation of probability densities, see Sec.~\ref{sec2.2}. The result~\eqref{joint-2} agrees with the results derived in Refs.~\cite{AkemannBur2012,Adhikari2013}.

For quaternion matrices ($\beta=4$) everything works along the same lines as for $\beta=2$. We combine Eqs.~\eqref{jpdfgen-bet4} and \eqref{measureGaus1} and  find
\begin{multline}\label{joint-4}
p_{{ \nu}}^{ ({\rm \gauss}, \beta=4)}(Z^{(M)})\propto \Delta_{2N}(Z^{(M)},{Z^{(M)}}^*)\\
\times\prod_{a=1}^{N_{\min}}(z_a-z_a^*)g^{({\rm \gauss}, M)}_{{ 2\nu}}(2^{M/2}z_a).
\end{multline}
The one-point weight $g^{({\rm \gauss},M)}_{{ 2\nu}}$ is  apart from a replacement $\nu\to2\nu$ exactly the same weight as for complex matrices, see Eq.~\eqref{weight-2b}.  This joint probability density was studied in Ref.~\cite{Ipsen2013}.

Let $N_{\min}=2\widetilde{N}+\chi$. The joint probability density for the real matrices is much more involved. Again the measures $\widehat{P}_j$ are deformed Gaussians, see Eq.~\eqref{measureGaus1}, but their arguments are now $2\times 2$ real random matrices instead of complex random variables, cf. Eqs.~\eqref{jpdfgen-bet1.ev.2} and \eqref{jpdfgen-bet1.odd.2}. Thus the joint probability density is
\begin{align}
& p_{{ \nu}}^{ ({\rm \gauss},  \beta=1)}(Z^{(M)})\propto |\Delta_{2\widetilde{N}}(Z^{(M)})| \nn\\
&\times\prod_{a=1}^{\widetilde{N}}\left(\delta(\IM\,z_{2a-1})\delta(\IM\,z_{2a})+2\delta^{(2)}(z_{2a-1}-z_{2a}^*)\right)\nn\\
&\times h^{({\rm \gauss}, M)}_{{ \nu}}(z_{2a-1},z_{2a})\label{joint-1eva}
\end{align}
for even matrix dimension and
\begin{align}
&p_{{ \nu}}^{( {\rm \gauss},\beta=1)}(Z^{(M)})\propto |\Delta_{2\widetilde{N}+1}(Z^{(M)})| \nn\\
&\times g^{({\rm \gauss}, M)}_{{ \nu/2}}(\RE\,z_{2\widetilde{N}+1})\delta(\IM\,z_{2\widetilde{N}+1}) \nn\\
&\times\prod_{a=1}^{\widetilde{N}}\left(\delta(\IM\,z_{2a-1})\delta(\IM\,z_{2a})+2\delta^{(2)}(z_{2a-1}-z_{2a}^*)\right)\nn\\
&\times h^{({\rm \gauss}, M)}_{{ \nu}}(z_{2a-1},z_{2a})\label{joint-1odda}
\end{align}
for odd dimension.
The one-point weight, $g^{(M)}_{{ (\nu-1)/2}}$, is again the Meijer $G$-function~\eqref{weight-2b} but now with the indices ${ \nu/2}=(\nu_1/2,\ldots,\nu_M/2)$. It becomes a Gaussian in the Ginibre case ($M=1$, see Ref.~\cite{Sommers2008}) and the Bessel function of the second kind in the chiral Ginibre case ($M=2$, see Ref.~\cite{AkemannPhi2010}). The two point weight is
\begin{align}\label{two-point-weight}
& h^{ ({\rm \gauss}, M)}_{{ \nu}}(z_1,z_2)=\\
&\int_{{|\IM(z_{1}-z_{2})|/2}}^\infty \hspace*{-3em}d\alpha\hspace*{2em} \delta^{(4)}(\Lambda(z_{1},z_{2},\alpha)-\widehat{Z}^{(M)}\cdots \widehat{Z}^{(1)})\nn\\
&\times\prod_{j=1}^{M}\int d[\widehat{Z}^{(j)}] \, |{\det} \widehat{Z}^{(j)}|^{\nu_j} \exp[-\tr\widehat{Z}^{(j)}\widehat{Z}^{(j)\,T}].\nn
\end{align}

\begin{widetext}
This integral can be partially performed by first substituting $Y_1=\widehat{Z}^{(1)}$ and $Y_j=\widehat{Z}^{(j)}Y_{j-1}$ and then evaluating the four dimensional Dirac $\delta$-function such that
\begin{multline}\label{two-point-weight.2}
  h^{ ({\rm \gauss}, M)}_{{ \nu}}(z_1,z_2)= |z_{1}z_{2}|^{\nu_M}\int_{|\IM(z_{1}-z_{2})|/2}^\infty d\alpha\left(\prod_{j=1}^{M-1}\int d[Y^{(j)}]  |{\det} Y^{(j)}|^{\nu_j-\nu_{j+1}-2} \right)\\
 \times\exp\left[-\tr \Lambda^2(z_{1},z_{2},\alpha)(Y_{M-1}^TY_{M-1})^{-1}-\sum_{i=2}^{M-1}\tr Y_i^{T}Y_i(Y_{i-1}^TY_{i-1})^{-1}-\tr Y_1^{T}Y_1\right].
\end{multline}
Performing singular value decompositions for each of the matrices $Y_j$ and integrating over the corresponding groups yields
\begin{multline}\label{two-point-weight.3}
  h^{ ({\rm \gauss}, M)}_{{ \nu}}(z_1,z_2)\propto  |z_{1}z_{2}|^{\nu_M}\int_{|\IM(z_{1}-z_{2})|/2}^\infty \hspace*{-3em} d\alpha \hspace*{2em}
\bigg(\prod_{j=1}^{M-1}\int_0^\infty dy_{1j}\int_0^\infty dy_{2j} |y_{1j}^2-y_{2j}^2| |y_{1j} y_{2j} |^{\nu_j-\nu_{j+1}-2} \bigg)\\
\times\exp\left[-\frac{(4\alpha^2+z_1^2+z_2^2)(y_{1M-1}^2+y_{2M-1}^2)}{2y_{1M-1}^2y_{2M-1}^2}-\sum_{i=2}^{M-1}\frac{(y_{1i}^2+y_{2i}^2)(y_{1i-1}^2+y_{2i-1}^2)}{2y_{1i-1}^2y_{2i-1}^2}-y_{11}^2-y_{21}^2\right]\\
\times I_0\left(\frac{\sqrt{4\alpha^2+(z_1+z_2)^2}\sqrt{4\alpha^2+(z_1-z_2)^2}(y_{2M-1}^2-y_{1M-1}^2)}{2y_{1M-1}^2y_{2M-1}^2}\right)
\prod_{i=2}^{M-1}I_0\left(\frac{(y_{1i}^2-y_{2i}^2)(y_{2i-1}^2-y_{1i-1}^2)}{2y_{1i-1}^2y_{2i-1}^2}\right),
\end{multline}
where $I_0$ is the modified Bessel function of the first kind. For the Ginibre ensemble, i.e. $M=1$ and $\nu_1=0$, we can easily deduce the  error function in the imaginary part of the complex eigenvalue pair as it was found in Ref.~\cite{Sommers2008}. The case $M=2$, $\nu_1=0$ and $\nu_2=\nu$ arbitrary is the chiral Ginibre ensemble discussed in Refs.~\cite{AkemannPhi2010}. For arbitrary $M$ the integral~\eqref{two-point-weight.3} is a generalization of these two particular cases.
\end{widetext}

Notice that in the case of a real pair of eigenvalues the two-point weight reduces to a product of one point weights,
\begin{multline}\label{two-point-weight.4}
 h^{ ({\rm \gauss}, M)}_{{ \nu}}(\RE\,z_1,\RE\,z_2)\propto
 g^{({\rm \gauss}, M)}_{{ \nu/2}}(\RE\,z_1)g^{({\rm \gauss}, M)}_{{ \nu/2}}(\RE\,z_2).
\end{multline}
Although this is not immediately clear from the integral~\eqref{two-point-weight.3} it can be derived by a generalized Schur decomposition of the $2\times2$ blocks, see Ref.~\cite{Forrester2013}.

\subsection{Products of Jacobi matrices}\label{sec3.3}

Here we consider random matrices drawn from  Jacobi ensembles, where the integrals over the strictly upper triangular matrices $\Delta^{(j)}$ are more involved than in the Gaussian case. Let us briefly discuss how to perform these integrations for $\beta=2$. The derivation for $\beta=1,4$ works in a similar way. Starting with the Jacobi measure~\eqref{measuretrunc1}, we perform a generalized Schur decomposition decomposition for the individual matrices,
\begin{equation}
\widetilde X_j=U_j^{-1}S^{(j)}U_{j-1}\quad\text{with}\quad S^{(j)}=Z_j+\Delta^{(j)}.
\end{equation}
As usual the $Z_j$'s denote the diagonal matrices, while $\Delta^{(j)}$ are strictly upper triangular matrices. The unitary matrices, $U_k$, are trivially absorbed due to the invariance of the measures. We want to integrate over $\Delta^{(j)}$ in Eq.~\eqref{neww-bet2},
\begin{multline}\label{3.3.1}
\widehat{P}_j(Z_j)\propto\int {\det}^{\kappa_j+\nu_{j-1}}(\eins_{N_{\min}}-S^{(j)}S^{(j)\,\dagger})\\
\times{\det}^{\nu_{j}}(S^{(j)}S^{(j)\,\dagger})\Theta(\eins_{N_{\min}}-S^{(j)}S^{(j)\,\dagger})d[\Delta^{(j)}].
\end{multline}
 Notice that the second determinant can be pushed out the integral since it only depends on $Z_j$. In the first step we split the $N_{\rm min}\times N_{\rm min}$ upper triangular matrix $S^{(j)}$ like
\begin{equation}\label{split}
 S^{(j)}=\left[\begin{array}{cc} {S'}^{(j)} & v^{(j)} \\ 0 & z_{N_{\rm min}}^{(j)} \end{array}\right],
\end{equation}
where ${S'}^{(j)}$ is a $(N_{\rm min}-1)\times (N_{\rm min}-1)$ upper triangular matrix and $v^{(j)}$ a $(N_{\rm min}-1)$-dimensional vector. Thus we have
\begin{multline}
  \det(\eins_{N_{\min}}-S^{(j)}S^{(j)\,\dagger})=\\
\det\!\left[1-|z_{N_{\min}}^{(j)}|^2-v^{(j)\,\dagger}\!\left(\eins_{N_{\min}-1}-{S'}^{(j)}{S'}^{(j)\,\dagger}\right)^{-1}\!\!\!v^{(j)}\right]\\
\times\det\left(\eins_{N_{\min}-1}-{S'}^{(j)}{S'}^{(j)\,\dagger}\right).\label{3.3.2}
\end{multline}
Rescaling
\begin{equation}
v^{(j)}\rightarrow \sqrt{(1-|z_{N_{\min}}^{(j)}|^2)\left(\eins_{N_{\min}-1}-{S'}^{(j)}{S'}^{(j)\,\dagger}\right)}v^{(j)}
\end{equation}
the integral over $v^{(j)}$ factorizes and yields a constant such that we get
\begin{align}
\widehat{P}_j(Z_j)&\propto(1-|z_{N_{\min}}^{(j)}|^2)^{\kappa_j+N_{j-1}-1}\Theta(1-|z_{N_{\min}}^{(j)}|^2)\nn\\
&\times|{\det}Z^{(j)}|^{2\nu_{j}}\int d[{\Delta'}^{(j)}]\Theta(\eins_{N_{\min}-1}-{S'}^{(j)}{S'}^{(j)\,\dagger})\nn\\
&\times{\det}^{\kappa_j+\nu_{j-1}+1}(\eins_{N_{\min}-1}-{S'}^{(j)}{S'}^{(j)\,\dagger}),
\label{3.3.3}
\end{align}
where ${\Delta'}^{(j)}$ is the strictly upper triangular part of ${S'}^{(j)}$. This procedure can be iterated and we find the well-known induced probability density \cite{Khoruzhenko2010,Forresterbook,Adhikari2013,AkemannKieBur2013}
\begin{multline}\label{prob-den-bet2}
\widehat{P}_j(Z_j)\propto|{\det}Z^{(j)}|^{2\nu_{j}}
{\det}^{\kappa_j+N_{j-1}-1}(\eins_{N_{\min}}-|Z^{(j)}|^2) \\
\times\Theta(\eins_{N_{\min}}-|Z^{(j)}|^2)
\end{multline}
for $\beta=2$. In the real and quaternion case one can readily extend this procedure and finds the induced probability densities
\begin{multline}\label{prob-den-bet4}
  \widehat{P}_j(Z_j)\propto|{\det}Z^{(j)}|^{4\nu_{j}}{\det}^{2(\kappa_j+N_{j-1}-1)}(\eins_{N_{\min}}-|Z^{(j)}|^2) \\
\times\Theta(\eins_{N_{\min}}-|Z^{(j)}|^2)
\end{multline}
for $\beta=4$ and
\begin{multline}\label{prob-den-bet1-even}
  \widehat{P}_j(\widehat{Z}_j)\propto|{\det}\widehat{Z}^{(j)}|^{\nu_{j}}{\det}^{\kappa_j+N_{j-1}/2-1}(\eins_{N_{\min}}-\widehat{Z}^{(j)}\widehat{Z}^{(j)\,T}) \\
\times\Theta(\eins_{N_{\min}}-\widehat{Z}^{(j)}\widehat{Z}^{(j)\,T})
\end{multline}
for $\beta=1$ and even $N_{\min}$ and
\begin{multline}
  \widehat{P}_j(\widehat{Z}_j)\propto|{\det}\widehat{Z}^{(j)}|^{\nu_{j}}{\det}^{\kappa_j+N_{j-1}/2-1}(\eins_{N_{\min}}-\widehat{Z}^{(j)}\widehat{Z}^{(j)\,T}) \\
 \times \sqrt{1-\widehat{z}_{N_{\min}}^{(j)}}  \Theta(\eins_{N_{\min}}-\widehat{Z}^{(j)}\widehat{Z}^{(j)\,T})\label{prob-den-bet1-odd}
\end{multline}
for odd $N_{\min}$. Recall that we have a block diagonal structure of $\widehat{Z}$ consisting of $2\times 2$ blocks in the real case.

The joint probability density of the product matrix $X^{(M)}$ can be readily read off for $\beta=2,4$ and is
\begin{equation}\label{joint-trunc-2}
  p_{{ \nu,\mu}}^{( {\rm \jacobi}, \beta=2)}(Z^{(M)})\propto |\Delta_{N_{\min}}(Z^{(M)})|^2\prod_{a=1}^{N_{\min}}g^{({\rm \jacobi}, M)}_{{ \nu,\mu}}(z_a)
\end{equation}
for $\beta=2$, cf. Refs.~\cite{Adhikari2013,AkemannKieBur2013}, and
\begin{multline}\label{joint-trunc-4}
  p_{{ \nu,\mu}}^{( {\rm \jacobi}, \beta=4)}(Z^{(M)})\propto \Delta_{2N_{\min}}(Z^{(M)},Z^{(M)\,*})\\
\times\prod_{a=1}^{N_{\min}}(z_a-z_a^*)g^{({\rm \jacobi}, M)}_{{ 2\nu,2\mu-1}}(z_a)
\end{multline}
for $\beta=4$. The one-point weight is this time
\begin{multline}
  g^{( {\rm \jacobi}, M)}_{{ \nu,\mu}}(z)=
\prod_{j=1}^{M}\int_{|z^{(j)}|=1} \hspace*{-2em} d^2 z^{(j)}\ \frac{|z^{(j)}|^{2\nu_j}(1-|z^{(j)}|^2)^{\mu_j-\nu_j-1}}{\Gamma(\mu_j-\nu_j)}\\
\times\delta^{(2)}(z-z^{(M)}\cdots z^{(1)}),
\label{weight-trunc-2a}
\end{multline}
where $\nu$ and $\mu$ collectively denote the constants $\nu_i=N_i-N_{\min}$ and $\mu_i=\kappa_i+\nu_i+N_{i-1}$, respectively. Recall that the ordering of the indices is irrelevant due to the weak communication relation.
The one point weight can be again expressed as a Meijer $G$-function \cite{Gradshteynbook},
\begin{align}\label{weight-trunc-2b}
 g^{( {\rm \jacobi}, M)}_{{ \nu,\mu}}(z)&= \MeijerG{M}{0}{M}{M}{\mu_1,\ldots,\mu_M}{\nu_1,\ldots,\nu_M}{|z|^2}\\
 &=\int_{\mathcal{C}}\frac{du}{2\pi\imath}|z|^{2u}\prod_{j=1}^M\frac{\Gamma(\nu_j-u)}{\Gamma(\mu_j-u)}.\nonumber
\end{align}
 
The analogue of the joint probability densities~\eqref{joint-1eva} and \eqref{joint-1odda} for a product of truncated orthogonal matrices is
\begin{align}
&   p_{{ \nu}}^{ ({\rm \jacobi},  \beta=1)}(Z^{(M)})\propto |\Delta_{2\widetilde{N}}(Z^{(M)})|\nn \\
&\times\prod_{a=1}^{\widetilde{N}}\left(\delta(\IM\,z_{2a-1})\delta(\IM\,z_{2a})+2\delta^{(2)}(z_{2a-1}-z_{2a}^*)\right)\nn \\
&\times h^{({\rm \jacobi}, M)}_{{ \nu}}(z_{2a-1},z_{2a})\label{joint-1eva-trunc}
\end{align}
for an even matrix dimension and
\begin{align}
&   p_{{ \nu}}^{( {\rm \jacobi},\beta=1)}(Z^{(M)})\propto|\Delta_{2\widetilde{N}+1}(Z^{(M)})|\nn\\
&\times g^{({\rm \jacobi}, M)}_{{ \nu/2,\tilde{\mu}}}(\RE\,z_{2\widetilde{N}+1})\delta(\IM\,z_{2\widetilde{N}+1})\nn\\
&\times \prod_{a=1}^{\widetilde{N}}\left(\delta(\IM\,z_{2a-1})\delta(\IM\,z_{2a})+2\delta^{(2)}(z_{2a-1}-z_{2a}^*)\right)\nn\\
&\times h^{({\rm \jacobi}, M)}_{{ \nu}}(z_{2a-1},z_{2a})\label{joint-1odda-trunc}
\end{align}
for an odd dimension. Here $\tilde\mu$ collectively denotes the constants $\tilde{\mu}_i=\kappa_i+(\nu_i+N_i+1)/2$. The two-point weight is in this case
\begin{align}
& h^{ ({\rm \jacobi}, M)}_{{ \nu,\tilde\mu}}(z_1,z_2)=\nn\\
& \int_{|\IM(z_{1}-z_{2})|/2}^\infty \hspace*{-3em} d\alpha \hspace*{2em} \delta^{(4)}(\Lambda(z_{1},z_{2},\alpha)-\widehat{Z}^{(M)}\cdots \widehat{Z}^{(1)}) \nn\\
&\times\biggl[\prod_{j=1}^{M}\int d[\widehat{Z}^{(j)}] \, |{\det} \widehat{Z}^{(j)}|^{\nu_j} \Theta(\eins_{N_{\min}}-\widehat{Z}^{(j)}\widehat{Z}^{(j)\,T}) \nn\\
&\qquad\qquad\times {\det}^{\tilde{\mu}_j-(\nu_j+3)/2}(\eins_{N_{\min}}-\widehat{Z}^{(j)}\widehat{Z}^{(j)\,T})\biggl].\label{two-point-weight-trunc}
\end{align}
This weight can be also rephrased to something like Eq.~\eqref{two-point-weight.3} which we omit here since it looks quite complicated and does not yield new insights. Let us state, at least, what the weight for a real eigenvalue pair is
\begin{multline}
  h^{ ({\rm \jacobi}, M)}_{{ \nu},\tilde\mu}(\RE\,z_1,\RE\,z_2)\propto
 g^{({\rm \jacobi}, M)}_{{ \nu/2,\tilde{\mu}}}(\RE\,z_{1})g^{({\rm \jacobi}, M)}_{{ \nu/2,\tilde{\mu}}}(\RE\,z_{2}).\label{two-point-weight-trunc.2}
\end{multline}
Again this can be derived by performing a generalized Schur decomposition of the $2\times 2$ blocks along the idea of Ref.~\cite{Forrester2013}.

\section{Eigenvalue correlation functions and the Lyapunov exponent of the open, chaotic chain}\label{sec4}

In this section we derive the eigenvalue correlation functions of products of Ginibre matrices, Jacobi matrices and an intermix of both kinds. Furthermore, we discuss the Lyapunov exponents of the eigenvalues of the product matrices. From the structure of the joint probability densities discussed in the previous section, we can immediately conclude that all eigenvalue correlations can be reduced to averages over one and two characteristic polynomials, which are thus the fundamental objects and determine the whole eigenvalue statistics. Moreover we can conclude that the $k$-point correlation functions as well as the averages over an arbitrary number of ratios of characteristic polynomials follow determinantal ($\beta=2$) and Pfaffian ($\beta=1,4$) point processes. The reason is that the joint probability densities only depend on a product of a squared Vandermonde determinant and one-point weights ($\beta=2$) corresponding to bi-orthogonal polynomials or on a Vandermonde 
determinant and a product of two-point weights ($\beta=1,4$) corresponding to skew-orthogonal polynomials. There is a whole scope of literature discussing such ensembles, see Refs.~\cite{Mehtabook,Bergere2004,Borodin2005,Kieburg2009a,Kieburg2009b,Forresterbook,Oxfordbook} and references therein.

Indeed the determinantal and Pfaffian point processes carry over to a mixed product of Ginibre and Jacobi matrices due to the simple structure of both kinds of ensembles. This can be easily seen when considering the joint probability density of the eigenvalues of a product matrix $X^{(M_1+M_2)}=X_{M}X_{M-1}\cdots X_1$, where $X_j$, $j\in I_1=\{j_1,\ldots,j_{M_1}\}$, are complex Ginibre matrices and $X_i$, $i\in I_2=\{i_1,\ldots,i_{M_2}\}$, are truncated unitary matrices. The index sets $I_1$ and $I_2$ have an empty section, i.e. $I_1\cap I_2=\emptyset$, and a union equal to $I_1\cup I_2=\{1,\ldots, M_1+M_2=M\}$.

One can consider eigenvalues (in a generalized sense) of rectangular matrices~\cite{BSS:2007,WT:2002}, but here we restrict ourselves to square matrices, hence we choose $N_0=N_M$. Equivalently one can consider the induced product matrix $\widetilde X^{(M)}$, cf. Eq.~\eqref{newproduct}, which is a square matrix by definition. Then one can trivially combine the results~\eqref{joint-2} and \eqref{joint-trunc-2} and finds
\begin{equation}\label{joint-comb-2}
  p_{{ \nu,\mu}}^{(\beta=2)}(Z^{(M)})\propto
 |\Delta_{N_{\min}}(Z^{(M)})|^2\prod_{a=1}^{N_{\min}}g^{(M_1,M_2)}_{{ \nu,\mu}}(z_a)
\end{equation}
with the one point weight
\begin{multline}
g^{(M_1,M_2)}_{{ \nu,\mu}}(z)\propto
\prod_{j\in I_1}\int_{\mathbb{C}} d^2z^{(j)} |z^{(j)}|^{2\nu_j}\exp[-|z^{(j)}|^2]\\
\times\prod_{j\in I_2}\int_{|z^{(j)}|=1} \hspace*{-2em} d^2z^{(j)}\ \frac{|z^{(j)}|^{2\nu_j}(1-|z^{(j)}|^2)^{\mu_j-\nu_j-1}}{\Gamma(\mu_j-\nu_j)}\\
\times\delta^{(2)}(z-z^{(M)}\cdots z^{(1)}),
\label{weight-comb-2a}
\end{multline}
where $\nu$ and $\mu$ collectively denote the constants $\nu_i=N_i-N_{\min}$ and $\mu_{i_k}=\kappa_{i_k}+\nu_{i_k}+N_{i_k-1}$.
Again, the one-point weight can be expressed as a Meijer $G$-function~\cite{Gradshteynbook},
\begin{align}
g^{(M_1,M_2)}_{{ \nu,\mu}}(z)&= \MeijerG{M_1+M_2}{0}{M_2}{M_1+M_2}{\mu_{i_1},\ldots,\mu_{i_{M_2}}}{\nu_1,\ldots,\nu_{M_1+M_2}}{|z|^2}\nn\\
&=\int_{\mathcal{C}}\frac{du}{2\pi\imath}|z|^{2u}\frac{\prod_{j=1}^{M_1+M_2}\Gamma(\nu_j-u)}{\prod_{i\in I_2}\Gamma(\mu_i-u)}.
\label{weight-comb-2b}
\end{align}
Note that the weak commutation relation manifests itself in the weight through the invariance under permutations of the indices.
The special cases where the product consists solely of Ginibre or Jacobi matrices are deduced from this result by setting either $M_1$ or $M_2$ equal to zero. 

Similar results can be obtained for the case of real and quaternion matrices. Here we will only state the quaternion case ($\beta=4$),
\begin{multline}\label{joint-comb-4}
   p_{{ \nu,\mu}}^{(\beta=4)}(Z^{(M)})\propto \Delta_{2N_{\min}}(Z^{(M)},Z^{(M)\,*})\\
\times\prod_{a=1}^{N_{\min}}(z_a-z_a^*)g^{(M_1,M_2)}_{{ 2\nu,2\mu-1}}(2^{M_1/2}z_a),
\end{multline}
which have a structure closely related to the complex case~\eqref{joint-comb-2}.

In the ensuing two subsections we derive the eigenvalue densities of the complex and quaternion case. The discussion of the real case ($\beta=1$) will be postponed to forthcoming publications.  Moreover we will consider the more general case~\eqref{joint-comb-2} and \eqref{joint-comb-4} of a mixed product of Ginibre and Jacobi matrices.

\subsection{Complex matrices ($\beta=2$)}\label{sec4.1}

Looking at the joint probability density~\eqref{joint-comb-2} it is immediately clear that the corresponding orthogonal polynomials are the monomials $z^a$ and $z^{*\, b}$, since the one-point weight is invariant under rotation in the complex phase. These monomials have the normalization~\cite{Gradshteynbook}
\begin{align}
 \int_{\mathbb{C}} & |z|^{2a} g^{(M_1,M_2)}_{{ \nu,\mu}}(z)d^2z \nn\\
&=\pi\int_{0}^\infty r^{a+1} \MeijerG{M_1+M_2}{0}{M_2}{M_1+M_2}{\mu_{i_1},\ldots,\mu_{i_{M_2}}}{\nu_1,\ldots,\nu_{M_1+M_2}}{r}\frac{dr}{r}\nonumber\\
&=\pi\frac{\prod_{j=1}^{M_1+M_2}\Gamma(\nu_j+a+1)}{\prod_{i\in I_2}\Gamma(\mu_i+a+1)}\label{norm}
\end{align}
with respect to the weight $g^{(M_1,M_2)}_{{ \nu,\mu}}(z)$. Hence the joint probability density can be rewritten into the following determinantal structure,
\begin{equation}\label{joint-det}
 p_{{ \nu,\mu}}^{(\beta=2)}(Z^{(M)})=\frac{1}{N_{\min}!}\det\limits_{1\leq a,b\leq N_{\min}}\left[K^{(N_{\min})}(z_a,z_b^*)\right],
\end{equation}
see Refs.~\cite{Mehtabook,Bergere2004,Oxfordbook,Kieburg2009a} and references therein. The kernel is given by
\begin{multline}
  K^{(N_{\min})}(z_a,z_b^*)=
\frac{1}{\pi}\sqrt{g^{(M_1,M_2)}_{{ \nu,\mu}}(z_a)g^{(M_1,M_2)}_{{ \nu,\mu}}(z_b)} \\
\times\sum_{l=0}^{N_{\min}-1}\frac{\prod_{i\in I_2}\Gamma(\mu_i+l+1)}{\prod_{j=1}^{M_1+M_2}\Gamma(\nu_j+l+1)} z_a^l z_b^{*\,l}.
\label{kernel}
\end{multline}
It follows immediately that the level density is given by
\begin{multline}
  \rho^{(N_{\min})}(z)=\frac{1}{\pi}g^{(M_1,M_2)}_{{ \nu,\mu}}(z)\\
\times\sum_{l=0}^{N_{\min}-1}\frac{\prod_{i\in I_2}\Gamma(\mu_i+l+1)}{\prod_{j=1}^{M_1+M_2}\Gamma(\nu_j+l+1)} |z|^{2l},
\label{density}
\end{multline}
where the density inherits the isotropic structure from the one-point weight. The normalization is chosen such that the integration over the density yields the generic number of non-zero eigenvalues, i.e. $\int \rho^{(N_{\min})}(z)d^2z=N_{\min}$. If $N_0=N_M>N_{\min}>0$ then there are  $N_M-N_{\min}$ generic zero modes. They will be reflected as additional Dirac $\delta$-functions in the density~\eqref{density}.

The macroscopic limit, $N_{\min}\to\infty$ and $\hat{\mu}_i=\mu_i/N_{\min}$ and $\hat{\nu}_i=\nu_i/N_{\min}$ fixed, of the level density~\eqref{density} can be obtained by the scaling $\hat{z}= N_{\min}^{M_1/2} z$. Notice that we do not scale with $N_{\min}^{(M_1+M_2)/2}$ since the spectrum of those matrices drawn from a truncation of unitary matrices is of order one while the spectrum of the Ginibre matrices is of order $\sqrt{N_{\min}}$. Then the macroscopic level density is
\begin{equation}
 \rho(\hat{z})=\underset{N_{\min}\to\infty}{\lim}\frac{1}{N_{\min}^{M_1+1} }\rho^{(N_{\min})}\left(\frac{\hat{z}}{N_{\min}^{M_1/2} }\right).\label{macdensity}
\end{equation}
The easiest way to derive this level density is via the moments of this density
\begin{align}
 \langle |z|^{2k}\rangle_{\rho^{(N_{\min})}} &=\frac{1}{N_{\min}}\int_{\mathbb{C}}|z|^{2k}\rho^{(N_{\min})}(z)d\RE\,zd\IM\,z\nn\\
 &=\frac{1}{N_{\min}}\sum_{l=0}^{N_{\min}-1}\prod_{i\in I_2}\frac{\Gamma(\mu_i+l+1)}{\Gamma(\mu_i+l+k+1)}\nn\\
 &\qquad\times\prod_{j=1}^{M_1+M_2}\frac{\Gamma(\nu_j+l+k+1)}{\Gamma(\nu_j+l+1)}.\label{moments}
\end{align}
Employing the Stirling formula and approximating the sum by an integral we obtain
\begin{align}
 \langle |\hat{z}|^{2k}\rangle_{\rho} &=\underset{N_{\min}\to\infty}{\lim}\frac{1}{N_{\min}^{M_1k} }\langle |z|^{2k}\rangle_{\rho^{(N_{\min})}}\label{macmoments}\\
 &=\int_0^1\left(\frac{\prod_{j=1}^{M_1+M_2}(\hat{\nu}_j+y)}{\prod_{i\in I_2}(\hat{\mu}_i+y)}\right)^k dy.\nonumber
\end{align}
The macroscopic level density can be read off and it is
\begin{equation}
 \rho(\hat{z})=\frac{1}{\pi}\int_0^1\delta\left(R(y)-|\hat{z}|^2\right) dy\label{macdensity-2}
\end{equation}
with the rational function
\begin{equation}\label{ratiofunc}
R(y)=\frac{\prod_{j=1}^{M_1+M_2}(\hat{\nu}_j+y)}{\prod_{i\in I_2}(\hat{\mu}_i+y)}\geq 0,\quad\forall\, y\in[0,1].
\end{equation}

\begin{figure*}[htbp]
\includegraphics{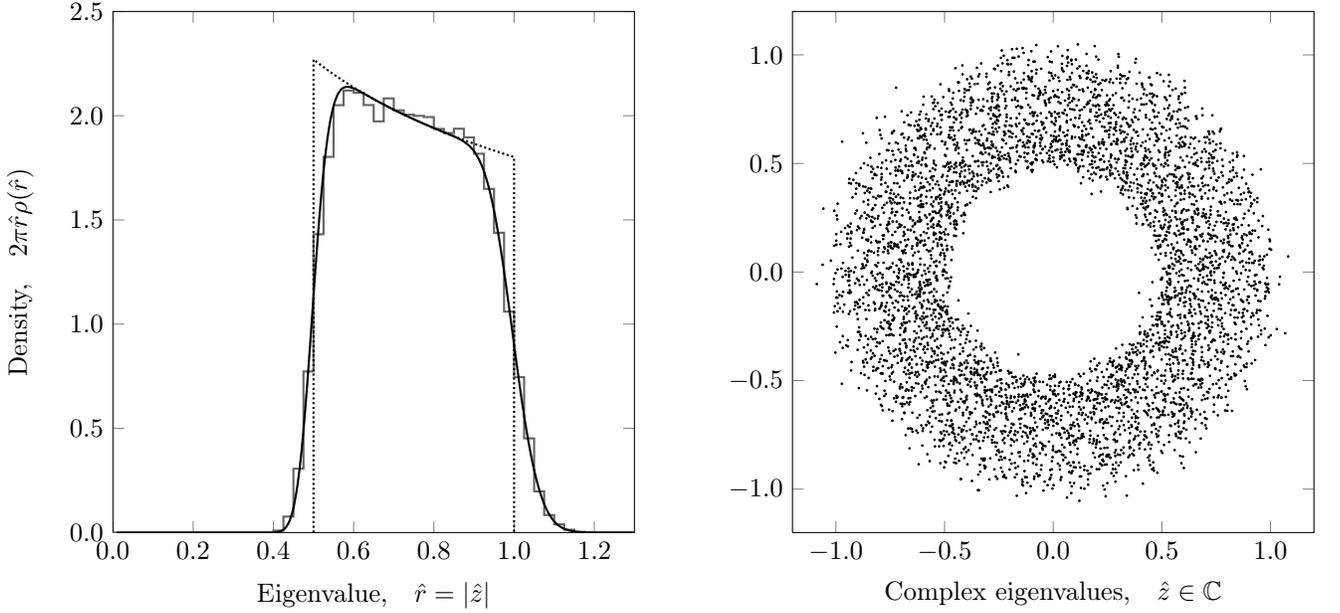}
\caption{Example of an induced product matrix, $\widetilde X^{(M)}$, with complex eigenvalues distributed within an annulus. The histogram depicted on the left panel shows the distribution of the absolute value of the eigenvalues for $500$ realizations of a product matrix with $N_{\min}=100 $, $M_1=3$, $M_2=0$ and $\hat\nu=\{1,2,3\}$; the solid curve shows the corresponding analytical prediction, while the dotted curve indicate the macroscopic limit. The right panel shows a scatter plot of $50$ out of the $500$ realizations generating the histogram on the left panel. Note that the fact that $\hat\nu_j>0$ for all $j$ implies that the original product matrix, $X^{(M)}=X_M\cdots X_1=U_L\widetilde X^{(M)}U_R$, is a rectangular matrix. }
\label{fig1}
\end{figure*}

Usually the domain of the level density $\rho(\hat{z})$ is a centred annulus in the complex plane~\cite{FeinbergZee,SingleRing}. To determine the inner and outer radius of the annulus it is quite convenient that $R(y)$ is strictly monotonous increasing on the interval $]0,1]$, i.e.
\begin{equation}
\frac{\partial}{\partial y}\ln R(y)=\sum_{j\in I_1}\frac{1}{\hat{\nu}_j+y}+\sum_{i\in I_2}\frac{\hat{\mu}_i-\hat{\nu}_i}{(\hat{\nu}_j+y)(\hat{\mu}_i+y)}>0.\nonumber\\
\label{derR}
\end{equation}
Notice that $\hat{\mu}_i>\hat{\nu}_i\geq0$ for all $i\in I_2$; compare $\kappa_i$ with the exponent of the determinant in Eq.~\eqref{Jacobimeas}. Therefore the inner and outer radius for the domain of $\rho(\hat{z})$ is
\begin{align}
r_{\min}&=R(0)=\frac{\prod_{j=1}^{M_1+M_2}\hat{\nu}_j}{\prod_{i\in I_2}\hat{\mu}_i}\quad\text{and} \\
r_{\max}&=R(1)=\frac{\prod_{j=1}^{M_1+M_2}(\hat{\nu}_j+1)}{\prod_{i\in I_2}(\hat{\mu}_i+1)},
\end{align}
respectively. Hence the inner radius vanishes if and only if one or more $\hat{\nu}_i$ vanish. If the inner radius vanishes the behaviour of the level density around the origin is $|\hat{z}|^{-2(\lambda-1)/\lambda}$ where $\lambda$ is the number of indices with $\hat{\nu}_i=0$. Note that if we are looking at a square product matrix $X^{(M)}=X_M\cdots X_1$, i.e. $N_0=N_M$, then it immediately follows that at least one $\hat\nu_j$ is equal to zero, and therefore that the inner radius vanishes such that the eigenvalues are located within a disk rather than an annulus. When starting from the induced product matrix, $\widetilde X^{(M)}=\widetilde X_M\cdots\widetilde X_1$, the level density can be still located within an annulus, see Fig.~\ref{fig1}. This mechanism is equivalent to that of induced Ginibre matrices~\cite{Fischmann2012}.

\begin{figure*}[htbp]
\includegraphics{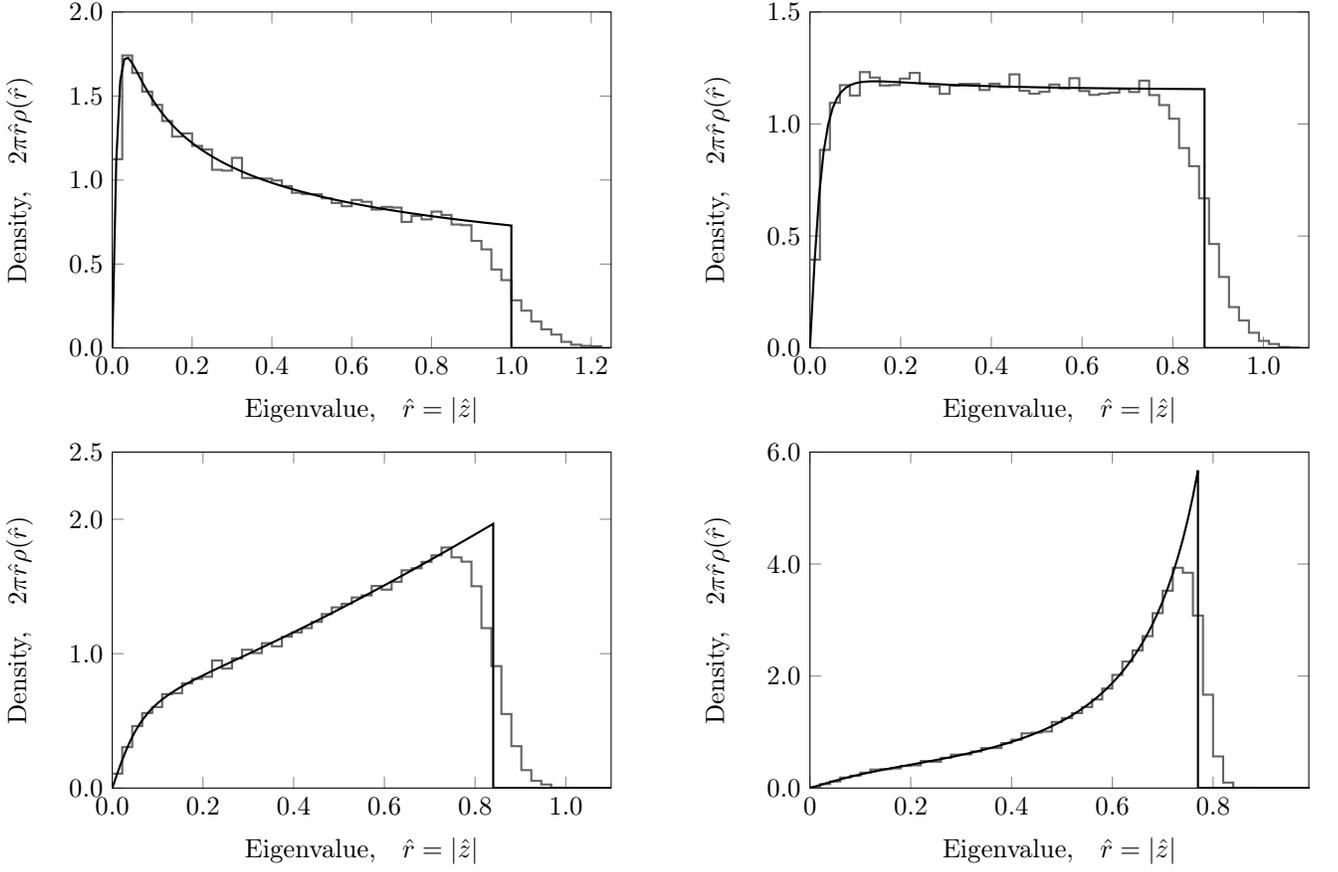}
\caption{Each histogram shows the distribution of the absolute value of the eigenvalues for $500$ realizations of a product of three independent complex ($\beta=2$) Ginibre and/or Jacobi random matrices with the smallest matrix dimension $N_{\min}=100$. The top left histogram has $M_1=3$, $M_2=0$ and $\hat\nu=\{0,1/10,2/10\}$, the top right histogram has $M_1=2$, $M_2=1$, $\hat\mu=\{3/10\}$ and $\hat\nu=\{0,1/10,2/10\}$, the bottom left histogram has $M_1=1$, $M_2=2$, $\hat\mu=\{3/10,3/10\}$ and $\hat\nu=\{0,1/10,2/10\}$, and the bottom right histogram has $M_1=0$, $M_2=3$, $\hat\mu=\{3/10,3/10,3/10\}$ and $\hat\nu=\{0,1/10,2/10\}$. The solid lines show the corresponding macroscopic limits, cf.~\eqref{macdensity-2}. Note that the axes on the four plots have different scales.}
\label{fig2}
\end{figure*}

For the Ginibre ensemble ($M_1=1$ and $M_2=0$) the density, $\rho(\hat{z})$, is the well-known complex unit disc with constant density~\cite{Mehtabook}. For general $M_1$ and $M_2=0$ the macroscopic level density was indirectly given by a polynomial equation of the Green function,
\begin{equation}\label{Greenfunc}
 G(z)=\int_\mathbb{C}\frac{\rho(\tilde{z}')d^2\tilde{z}}{\hat{z}-\tilde{z}}\ \Leftrightarrow\ \rho(z)=\frac{1}{\pi}\frac{\partial}{\partial \hat{z}^*}G(z),
\end{equation}
in Ref.~\cite{Burda2010-11}. Due to the isotropy of the level density the result of Ref.~\cite{Burda2010-11} can be readily deduced by the relation
\begin{equation}\label{invGreenfunc}
G(\hat{z})=\frac{1}{\hat{z}}\int_0^1\Theta\left(|\hat{z}|^2-R(y)\right) dy
=\frac{R^{-1}(|\hat{z}|^2)}{\hat{z}}
\end{equation}
($|\hat{z}|\leq R(1)$) yielding the polynomial equation ($M_2=0$)
\begin{equation}\label{polequ}
R(\hat{z}G(\hat{z}))=|\hat{z}|^2.
\end{equation}
The case of a truncated unitary matrix ($M_1=0$ and $M_2=1$) was discussed in Ref.~\cite{Zyczkowski2000}.

The macroscopic level density~\eqref{macdensity-2} can be easily numerically evaluated. The simplest way is to employ one of the many representations of the Dirac $\delta$-function as a limiting function. In Figs.~\ref{fig1} and \ref{fig2} we show the comparison of the macroscopic limit with numerical simulations for certain ensembles.

Let us return to the transport on the closed chain coupled to a particle bath, see Sec.~\ref{sec1}, one can easily calculate the average Lyapunov exponent ${\rm Lya}=\ln z=\ln r+\imath \varphi$, where we write the complex eigenvalues in polar coordinates, $z=re^{\imath\phi}$. Potential wells with a size comparable to the bath are modelled by Jacobi ensembles, while potential wells with a size much smaller than the bath are modelled by Ginibre ensembles.

Since the level density is isotropic the angular part vanishes while the radial part yields the mean Lyapunov exponent
\begin{equation}\label{Lyapunov-2a}
 \langle {\rm Lya}\rangle=2\pi\int_0^\infty \rho^{(N_{\min})}(r)\, r\ln r\,dr
\end{equation}
at finite matrix dimension. This integral simplifies in the large $N_{\min}$ limit  and we find
\begin{equation}\label{Lyapunov-2b}
 \langle{\rm Lya}\rangle=\frac{1}{2}\int_0^1\ln R(y)\, dy-\frac{M_1}{2}\ln N_{\min}.
\end{equation}
Interestingly the leading term, $\ln N_{\min}$, vanishes if the size of all potential wells is comparable to the bath, meaning that we have no random matrices drawn from the Ginibre ensemble. Any coupling of a Ginibre matrix with the bath implies that all particles will be sucked away after one round on the closed chain.

\subsection{Quaternion matrices ($\beta=4$)}\label{sec4.2}

We start from the joint probability density~\eqref{joint-comb-4} with the one-point weight~\eqref{weight-comb-2b}. Pursuing the calculation of the corresponding skew-orthogonal polynomials $p_j$ in Ref.~\cite{Ipsen2013}, i.e.
\begin{align}
 \left\langle p_a |p_b\right\rangle&=\int_{\mathbb{C}} d^2z (z^*-z)g^{(M_1,M_2)}_{{ 2\nu,2\mu-1}}(2^{M_1/2}z) \nn\\
&\qquad\qquad\times(p_a(z)p_b(z^*)-p_b(z)p_a(z^*))\nn\\
&=\left\{\begin{array}{r@{\quad}l} h_l, & a=2l+1, b=2l, \\ -h_l, & a=2l, b=2l+1, \\ 0, & \text{otherwise},\end{array}\right.
\label{skew-orth-def}
\end{align}
we find the polynomials
\begin{align}
  p_{2l}(z)&=\sum_{k=0}^l \left(\prod_{n=k+1}^l2^{-M_1}\frac{\prod_{j=1}^{M_1+M_2}(2\nu_j+2n)}{\prod_{i\in I_2}(2\mu_i+2n-1)}\right)z^{2k},\nn\\
  p_{2l+1}(z)&=z^{2l+1}\label{skew-orth-res}
\end{align}
and the normalization
\begin{equation}\label{norm-res}
 h_l=2\pi\  2^{-2M_1(l+1)}\frac{\prod_{j=1}^{M_1+M_2}\Gamma(2\nu_j+2l+2)}{\prod_{i\in I_2}\Gamma(2\mu_i+2l+1)}
\end{equation}
in agreement with Ref.~\cite{Ipsen2013} which was for $M_2=0$ and $\nu_j=\nu$ for all $j=1\ldots M_1$. This result directly follows from the isotropy of the one-point weight and the moments of this weight \cite{Gradshteynbook},
\begin{multline}
 \int_0^\infty \!\!\MeijerG{M_1+M_2}{0}{M_2}{M_1+M_2}{2\mu_{i_1}-1,\ldots,2\mu_{i_{M_2}}-1}{2\nu_1,\ldots,2\nu_{M_1+M_2}}{2^{M_1}r} r^{l} \frac{dr}{r} \\
=2^{-M_1l}\frac{\prod_{j=1}^{M_1+M_2}\Gamma(2\nu_j+l)}{\prod_{i\in I_2}\Gamma(2\mu_i+l-1)}.
 \label{moments4.2}
\end{multline}
Thus the joint probability density can be written as a Pfaffian, see Refs.~\cite{Mehtabook,Oxfordbook,Kieburg2009b} and references therein,
\begin{multline}\label{joint-Pfaff}
 p_{{ \nu,\mu}}^{(\beta=4)}(Z^{(M)})= \\
\frac{1}{N_{\min}!}\pf_{1\leq a,b\leq N_{\min}}\left[\begin{array}{cc} \widehat{K}^{(N_{\min})}(z_a,z_b) & \widehat{K}^{(N_{\min})}(z_a,z_b^*) \\ \widehat{K}^{(N_{\min})}(z_a^*,z_b) & \widehat{K}^{(N_{\min})}(z_a^*,z_b^*) \end{array}\right]\\
\times\prod_{j=1}^{N_{\min}}(z_j^*-z_j)g^{(M_1,M_2)}_{{ 2\nu,2\mu-1}}(2^{M_1/2}z_j)
\end{multline}
with the pre-kernel
\begin{equation}
 \widehat{K}^{(N_{\min})}(z_a,z_b)=
\sum_{l=0}^{N_{\min}-1}\frac{p_{2l+1}(z_a)p_{2l}(z_b)-p_{2l}(z_a)p_{2l+1}(z_b)}{h_l}.\label{kernelPf}
\end{equation}
From Eq.~\eqref{joint-Pfaff} one can easily read off the level density which is
\begin{equation}\label{density-Pfaff}
 \rho^{( N_{\min}, \beta=4)}(z)=
(z^*-z)g^{(M_1,M_2)}_{{ 2\nu,2\mu-1}}(2^{M_1/2}z)\widehat{K}^{(N_{\min})}(z,z^*).
\end{equation}
We have normalized the density to $N_{\min}$ again. Note that, despite the isotropic one-point weight, the level density~\eqref{density-Pfaff} is not rotational symmetric. The reason is that the eigenvalues come in complex conjugate pairs, which results in a repulsion from the real axis.

The radial projection of the level density,
\begin{equation}\label{radproj0}
\rho^{( N_{\min}, \beta=4)}_{\rm proj}(r)=r \int_{-\pi}^\pi \rho^{( N_{\min}, \beta=4)}(re^{\imath\varphi})d\varphi,
\end{equation}
is an interesting quantity in many situations. For instance, hole probabilities and overcrowding at the origin only depend on the radial distribution. Moreover, the radial distribution is often useful for comparisons with numerical simulations because of the drastically improved statistics. The integral~\eqref{radproj0} yields
\begin{multline}\label{radproj}
\rho^{( N_{\min}, \beta=4)}_{\rm proj}(r)=
2^{M_1/2+1}g^{(M_1,M_2)}_{{ 2\nu,2\mu-1}}(2^{M_1/2}r)\\
\times\sum_{l=0}^{N_{\min}-1}\frac{\prod_{i\in I_2}\Gamma(2\mu_i+2l+1)}{\prod_{j=1}^{M_1+M_2}\Gamma(2\nu_j+2l+2)}(2^{M_1/2}r)^{4l+3}
\end{multline}
and looks quite similar to the level density of $\beta=2$, cf. Eq.~\eqref{density}. Actually, it yields the same macroscopic limit,
\begin{equation}\label{limit}
 \lim_{N_{\min}\to\infty}\frac{1}{N_{\min}^{M_1/2+1}}\rho^{( N_{\min}, \beta=4)}_{\rm proj}\left(\frac{\hat{r}}{N_{\min}^{M_1/2}}\right)=2\pi\hat{r}\rho(\hat{r}),
\end{equation}
cf. Eq.~\eqref{macdensity-2}. Thus the real part of the Lyapunov exponent of the transport model discussed in Sec.~\ref{sec1} has to be the same as for the symmetry class $\beta=2$ in this particular limit. We emphasize that this is not true for finite $N_{\min}$.

\section{Conclusions}\label{sec5}

We studied some general properties of a product of $M$ independent rectangular random matrices for all three Dyson classes in a unifying way. The only assumption on the weights of the individual matrices are their invariance under left- and right-multiplication of orthogonal, unitary and unitary symplectic matrices, respectively. These weights are also known as isotropic weights~\cite{Burda2013}. In this general context, we  showed that a product of rectangular random matrices is equivalent to a product of square matrices with modified weights. More strikingly we proved that the individual matrices in the product matrix satisfy a weak commutation relation. This weak commutation relation tells us that a product of independently distributed square random matrices (they can also result from rectangular matrices) is independent of the order of the product when averaging over them. Note that this weak commutation relation has immediate consequences in physical systems. For the considered example of a closed one-
dimensional chaotic chain in an environment, the 
ordering of the potential wells is irrelevant as long as we do not consider cross correlations. The same applies to telecommunications where the permutation of consecutive scatterers does not change the spectrum of the channel matrix, see \cite{AkemannKieWei2013,AkemannKieIps2013}. We underline that the weak commutation relation holds at finite matrix dimension and, thus, generalizes a known result for the macroscopic limit of the product of isotropic distributed matrices~\cite{Burda2013}. A weak commutation relation for products of random matrices has previously been discussed in the context of disordered wires with obstacles~\cite{BM:1994}.

The weak commutation relation holds on the level of matrices and affects therefore many quantities of physical interest, such as eigen- and singular values, but also the eigenvectors. We focused on the spectral properties; especially the eigenvalue correlations. We derived the joint probability density functions for all three Dyson classes and for general weights. In particular, we showed that a product of Ginibre matrices, a product of Jacobi matrices or an intermix of both kinds of matrices yields a determinantal (for $\beta=2$) or Pfaffian (for $\beta=1,4$) point process as it is well-known for many other ensembles, see \cite{Mehtabook,Bergere2004,Oxfordbook,Kieburg2009a,Kieburg2009b} and references therein.
We derived a representation of the one- and two-point weights in terms of a product of random variables for $\beta=2,4$ and in terms of a product of $2\times2$ random matrices for $\beta=1$. For the one-point weight we explicitly integrated over the random variables and showed that they are equal to Meijer $G$-functions which were already shown for particular cases in~\cite{AkemannBur2012,Ipsen2013,Adhikari2013,AkemannKieBur2013}. The numerical simulations performed for the product matrices are in complete agreement with the analytical results.

We also considered the macroscopic limit of such an intermixing product of Ginibre and Jacobi matrices and derived an explicit representation of the level density in terms of a one-fold integral. This result agrees with the implicit polynomial equation derived for the corresponding Green function in \cite{Burda2010-11}. We saw that the macroscopic level density either lives on an annulus or a complex disc centered around the origin, which is in agreement with the single ring theorem~\cite{FeinbergZee,SingleRing}. After proper unfolding, universality should hold on a local scale as have been partially discussed in~\cite{AkemannBur2012,Ipsen2013,AkemannKieBur2013}.

Finally, we briefly discussed the relation between matrix products and a closed one-dimensional chaotic chain in an environment. In particular, we calculate the Lyapunov exponents. We concluded that the ordering of the potential wells  is irrelevant as long as we do not consider cross correlations, which directly follows from the weak commutation relation. Furthermore, we showed that if at least one potential well on the chain is small compared to the bath, then all particles disappear from the chain after a single revolution.

\paragraph*{Acknowledgements:}

We acknowledge support by
the International Graduate College IRTG 1132 ``Stochastic and Real World Models'' of the German Science Foundation DFG (J.R.I). Moreover we thank Gernot Akemann for fruitful discussions and Peter Forrester for sharing with us his preprint version of Ref.~\cite{Forrester2013}.

\appendix

\begin{widetext}

\section{Distributional relation of eigenvalues and singular values of a $2\times2$ real matrix}\label{app1}

The given problem is the following. We have a joint probability distribution of  the singular values $\Lambda=\diag(\lambda_1,\lambda_2)$ of a $2\times2$ real matrix $Z=V_{\rm L}\Lambda V_{\rm R}$ with $\lambda_1\geq\lambda_2$, $Q(\lambda_1+\lambda_2,\lambda_1\lambda_2)|\lambda_1^2-\lambda_2^2|$, and $V_{\rm L/R}\in\Ort(2)$ distributed by the Haar measure. Notice that we assume $Q$ as a function of the trace and the modulus of the determinant of $Z$. What is the joint probability density of the eigenvalues of $Z$? To solve this question we pursue an idea similar to the calculations done in Refs.~\cite{Sommers2008, AkemannPhi2010}. 

We start from the zeros of the characteristic polynomials
\begin{equation}\label{charpol}
\det(V_{\rm L}\Lambda V_{\rm R}-z\eins_2)=0.
\end{equation}
From this equation we notice that only one rotation angle (from now on denoted by $\varphi\in[0,\pi[$) parametrizing $V_{\rm L}$ and $V_{\rm R}$ plays a role. The other one cancels out. Let the sign of the determinant of $Z$ be
\begin{equation}\label{sign}
s={\rm sign} (\det Z)={\rm sign}( \det V_{\rm L}\Lambda V_{\rm R})={\rm sign} (\det V_{\rm L}){\rm sign} (\det V_{\rm R}).
\end{equation}
Then Eq.~\eqref{charpol} can be rewritten to
\begin{equation}\label{charpol.2}
0=\det\left(\frac{\lambda_1-s\lambda_2}{2}
\left[\begin{array}{cc} 1 & 0 \\ 0& -1 \end{array}\right]+\frac{\lambda_1+s\lambda_2}{2}\left[\begin{array}{cc} \cos\varphi & \sin\varphi \\ -\sin\varphi & \cos\varphi \end{array}\right]-z\eins_2\right)
=z^2-(\lambda_1+s\lambda_2)\cos\varphi\, z+s\lambda_1\lambda_2.
\end{equation}
Thus the eigenvalues are
\begin{equation}\label{eigv}
 z_{\pm}=\frac{\lambda_1+s\lambda_2}{2}\cos\varphi\pm\sqrt{\frac{(\lambda_1+s\lambda_2)^2}{4}\cos^2\varphi-s\lambda_1\lambda_2}
 =\frac{\lambda_1+s\lambda_2}{2}\cos\varphi\pm\sqrt{\frac{(\lambda_1-s\lambda_2)^2}{4}-\frac{(\lambda_1+s\lambda_2)^2}{4}\sin^2\varphi}.
\end{equation}
We can only find a complex conjugate pair if $s=+1$ meaning $\det Z>0$.

Let $\lambda_\pm=(\lambda_1\pm\lambda_2)/2$. The joint probability density of the eigenvalues of $Z$ is given by
\begin{align}
 \widehat{Q}(z_1,z_2)&=\frac{8}{\pi}\int_0^\infty d\lambda_+ \int_0^{\lambda_+} d\lambda_-\int_0^\pi d\varphi Q(2\lambda_+,\lambda_+^2-\lambda_-^2)\lambda_+\lambda_-\delta^{(2)}(z_1-z_+)\delta^{(2)}(z_2-z_-).\label{jointp1} \\
 &=\frac{2}{\pi}\sum_{s=\pm}\int_0^\infty d\lambda_+ \int_0^{\lambda_+} d\lambda_-\int_0^\pi d\varphi Q(2\lambda_+,\lambda_+^2-\lambda_-^2)\lambda_+\lambda_-\nn\\
 &\times\delta^{(2)}\left(\lambda_s\cos\varphi-\frac{z_1+z_2}{2}\right) \delta^{(2)}\left(\sqrt{\lambda_{-s}^2-\lambda_s^2\sin^2\varphi}-\frac{z_1-z_2}{2}\right)\nonumber\\
 &=\frac{4}{\pi}\delta\left(\IM\,(z_1+z_2)\right)\sum_{s=\pm}\Theta(s z_1z_2)\int_0^\infty d\lambda_+^2 \int_0^{\lambda_+^2} d\lambda_-^2 Q(2\lambda_+,\lambda_+^2-\lambda_-^2)\nonumber\\
 &\times\frac{\Theta\left(\lambda_s-\left|\RE\,(z_1+z_2)\right|/2\right)}{\sqrt{\lambda_s^2-\RE^2(z_1+z_2)/4}} \delta^{(2)}\left(\sqrt{\lambda_{-s}^2-\lambda_s^2+\frac{\RE^2(z_1+z_2)}{4}}-\frac{z_1-z_2}{2}\right)\nonumber\\
 &=\widehat{Q}_{{\rm r},-}(z_1,z_2)+\widehat{Q}_{{\rm r},+}(z_1,z_2)+\widehat{Q}_{{\rm c}}(z_1,z_2).\nonumber
\end{align}
This distribution splits into three terms.  For $s=-1$ we have a distribution, $\widehat{Q}_{{\rm r},-}$, of two real eigenvalues where one is positive and the other one negative. For $s=+1$ we find a distribution, $\widehat{Q}_{{\rm r},+}$, of two real eigenvalues which are both positive or both negative (when $\lambda_{-}^2-\lambda_+^2+\RE^2(z_1+z_2)/4>0$) as well as one distribution, $\widehat{Q}_{{\rm c}}$, of a complex conjugate pair (only when $\lambda_{-}^2-\lambda_+^2+\RE^2(z_1+z_2)/4<0$).

First we concentrate on the case $s=-1$. We calculate
\begin{align}\label{jointp2}
 \widehat{Q}_{{\rm r},-}(z_1,z_2)&=\frac{4}{\pi}\delta\left(\IM\,z_1\right)\delta\left(\IM\,z_2\right)\Theta(- \RE\,z_1\RE\,z_2)\int_0^\infty d\lambda_+^2 \int_0^{\lambda_+^2} d\lambda_-^2 Q(2\lambda_+,\lambda_+^2-\lambda_-^2)\\
 &\times\frac{\Theta\left(\lambda_--\left|\RE\,(z_1+z_2)\right|/2\right)}{\sqrt{\lambda_-^2-\RE^2(z_1+z_2)/4}}\delta\left(\sqrt{\lambda_+^2-\lambda_-^2+\frac{\RE^2(z_1+z_2)}{4}}-\frac{\RE\,(z_1-z_2)}{2}\right)\nonumber\\
 &\overset{(1)}{=}\frac{4|\RE\,(z_1-z_2)|}{\pi}\delta\left(\IM\,z_1\right)\delta\left(\IM\,z_2\right)\Theta(- \RE\,z_1\RE\,z_2)
      \int_0^\infty d\lambda_+^2 \int_0^{1} d\widehat{\alpha} Q(2\lambda_+,\lambda_+^2(1-\widehat{\alpha}))\nonumber\\
 &\times\lambda_+^2\frac{\Theta\left(\lambda_+^2\widehat{\alpha}-\RE^2(z_1+z_2)/4\right)}{\sqrt{\lambda_+^2\widehat{\alpha}-\RE^2(z_1+z_2)/4}}\delta\left(\lambda_{+}^2(1-\widehat{\alpha})+\RE\,z_1\RE\,z_2\right)
      \delta\left(\lambda_{+}^2(\widehat{\alpha}-1)+\RE\,z_1\RE\,z_2\right)\nonumber\\
 &\overset{(2)}{=}\frac{4|\RE\,(z_1-z_2)|}{\pi}\delta\left(\IM\,z_1\right)\delta\left(\IM\,z_2\right)\Theta(- \RE\,z_1\RE\,z_2)
      \int_{\RE^2(z_1-z_2)/4}^\infty d\lambda_+^2 \frac{Q\left(2\lambda_+,-\RE\,z_1\RE\,z_2\right)}{\sqrt{\lambda_+^2-\RE^2(z_1-z_2)/4}}\nonumber\\
 &\overset{(3)}{=}\frac{8|z_1-z_2|}{\pi}\delta\left(\IM\,z_1\right)\delta\left(\IM\,z_2\right)\Theta(- \RE\,z_1\RE\,z_2)
      \int_0^\infty d\alpha Q\left(\sqrt{\RE^2(z_1-z_2)+4\alpha^2},-\RE\,z_1\RE\,z_2\right)\nonumber
\end{align}
We made the substitutions $\lambda_-=\lambda_+\sqrt{\widehat{\alpha}}$ in line (1) and  $\alpha=\sqrt{\lambda_+^2-\RE^2(z_1-z_2)/4}$ in line (3). In line (2) we integrated over $\widehat{\alpha}$ . A similar calculation can be also performed in the case of real eigenvalues with $s=+1$,
\begin{align}\label{jointp3}
 \widehat{Q}_{{\rm r},+}(z_1,z_2)&=\frac{4}{\pi}\delta\left(\IM\,z_1\right)\delta\left(\IM\,z_2\right)\Theta( \RE\,z_1\RE\,z_2)\int_0^\infty d\lambda_+^2 \int_0^{\lambda_+^2} d\lambda_-^2 Q(2\lambda_+,\lambda_+^2-\lambda_-^2)\\
 &\times\frac{\Theta\left(\lambda_+-\left|\RE\,(z_1+z_2)\right|/2\right)}{\sqrt{\lambda_+^2-\RE^2(z_1+z_2)/4}}\Theta\left(\lambda_{-}^2-\lambda_+^2+\frac{\RE^2(z_1+z_2)}{4}\right)
      \delta\left(\sqrt{\lambda_{-}^2-\lambda_+^2+\frac{\RE^2(z_1+z_2)}{4}}-\frac{\RE\,(z_1-z_2)}{2}\right)\nonumber\\
 &\overset{(1)}{=}\frac{4|\RE\,(z_1-z_2)|}{\pi}\delta\left(\IM\,z_1\right)\delta\left(\IM\,z_2\right)\Theta( \RE\,z_1\RE\,z_2)\nonumber\\
 &\times\int_0^\infty d\lambda_+^2 \int_0^{1} d\widehat{\alpha} Q(2\lambda_+,\lambda_+^2(1-\widehat{\alpha}))\lambda_+^2\frac{\Theta\left(\lambda_+-|\RE(z_1+z_2)|/2\right)}{\sqrt{\lambda_+^2-\RE^2(z_1+z_2)/4}}
      \delta\left(\lambda_{+}^2(\widehat{\alpha}-1)+\RE\,z_1\RE\,z_2\right)\nonumber\\
 &\overset{(2)}{=}\frac{4|\RE\,(z_1-z_2)|}{\pi}\delta\left(\IM\,z_1\right)\delta\left(\IM\,z_2\right)\Theta( \RE\,z_1\RE\,z_2)
      \times\int_{\RE^2(z_1+z_2)/4}^\infty d\lambda_+^2 \frac{Q\left(2\lambda_+,\RE\,z_1\RE\,z_2\right)}{\sqrt{\lambda_+^2-\RE^2(z_1+z_2)/4}}\nonumber\\
 &\overset{(3)}{=}\frac{8|z_1-z_2|}{\pi}\delta\left(\IM\,z_1\right)\delta\left(\IM\,z_2\right)\Theta( \RE\,z_1\RE\,z_2)
      \int_0^\infty d\alpha Q\left(\sqrt{\RE^2(z_1+z_2)+4\alpha^2},\RE\,z_1\RE\,z_2\right)\nonumber
\end{align}
Again we made some substitutions namely $\lambda_-=\lambda_+\sqrt{\widehat{\alpha}}$ in line (1) and $\alpha=\sqrt{\lambda_+^2-\RE^2(z_1+z_2)/4}$ in line (3). Step (2) is the same as in Eq.~\eqref{jointp2}. Please notice that this result is almost the same as in the case $s=-1$.

For the complex conjugated pair we have
\begin{align}\label{jointp4}
 \widehat{Q}_{\rm c}(z_1,z_2)&=\frac{8}{\pi}\delta^{(2)}\left(z_1-z_2^*\right)\int_0^\infty d\lambda_+^2 \int_0^{\lambda_+^2} d\lambda_-^2 Q(2\lambda_+,\lambda_+^2-\lambda_-^2)\\
 &\times\frac{\Theta\left(\lambda_+-\left|\RE\,z_1\right|\right)}{\sqrt{\lambda_+^2-\RE^2 z_1}}\Theta\left(\lambda_+^2-\lambda_-^2-\RE^2 z_1\right)\delta\left(\sqrt{\lambda_+^2-\lambda_-^2-\RE^2 z_1}-\IM\,z_1\right)\nonumber\\
 &\overset{(1)}{=}\frac{16|\IM\, z_1|}{\pi}\delta^{(2)}\left(z_1-z_2^*\right)\int_0^\infty d\lambda_+^2 \int_0^{1} d\widehat{\alpha} Q(2\lambda_+,\lambda_+^2(1-\widehat{\alpha}))
      \lambda_+^2\frac{\Theta\left(\lambda_+^2-\RE^2z_1\right)}{\sqrt{\lambda_+^2-\RE^2 z_1}}\delta\left(\lambda_+^2(1-\widehat{\alpha})-|z_1|^2\right)\nonumber\\
 &\overset{(2)}{=}\frac{16|\IM\, z_1|}{\pi}\delta^{(2)}\left(z_1-z_2^*\right)\int_{|z_1|^2}^\infty d\lambda_+^2 \frac{Q(2\lambda_+,|z_1|^2)}{\sqrt{\lambda_+^2-\RE^2 z_1}}\nonumber\\
 &\overset{(3)}{=}\frac{16|z_1-z_2|}{\pi}\delta^{(2)}\left(z_1-z_2^*\right)\int_{|\IM\,z_1|}^\infty d\alpha\,  Q(2\sqrt{\RE^2z_1+\alpha^2},|z_1|^2)\nonumber
\end{align}
The steps (1) and (3) are the same as in Eq.~\eqref{jointp3}. In the step (2) we recognize the fact that the Dirac $\delta$-function yields only something non-vanishing if $\lambda_+\geq |z_1|\geq|\RE\,z_1|\geq0$.

Comparing the results of Eqs.~(\ref{jointp2}-\ref{jointp4}) we notice that the function $Q$ always depends on $\sqrt{(z_1+\sign(z_1z_2) z_2)^2+4 \alpha}$ which is the argument for the trace of the original $2\times2$ matrix $Z$ and on $\sign(z_1z_2) z_1z_2$ which is the determinant of $Z$. If we want to rewrite the function $Q$ as a function of its singular values $\lambda_{1/2}$ we need the functional dependence of those variables on $z_{1/2}$ and $\alpha$. This dependence is
\begin{equation}\label{app.sing}
 \lambda_{1/2}=\left|\sqrt{\frac{(z_1+z_2)^2}{4}+\alpha^2}\pm\sqrt{\frac{(z_1-z_2)^2}{4}+\alpha^2}\right|.
\end{equation}
This relation readily follows from the system of equations,
\begin{equation}\label{syseq}
 \lambda_1+\lambda_2=\sqrt{(z_1+\sign(z_1z_2) z_2)^2+4 \alpha}\ {\rm and}\ \lambda_1\lambda_2=\sign(z_1z_2) z_1z_2,
\end{equation}
representing the trace and the determinant of $Z$, respectively. Again we emphasize that $z_1$ and $z_2$ are either real or complex conjugate.

\end{widetext}

\raggedright

\end{document}